\documentclass[preprint,12pt,number]{elsarticle}

\usepackage{amssymb}
\usepackage{amsmath}
\usepackage{siunitx}
\usepackage{url}
\usepackage{xurl}
\usepackage{booktabs}
\usepackage{pdflscape}
\usepackage{subcaption}
\usepackage[a4paper,margin=2.2cm]{geometry}
\usepackage{tabularx}
\usepackage{tikz}
\usetikzlibrary{arrows.meta,positioning}
\definecolor{rmInput}{RGB}{112,48,160}
\definecolor{rmData}{RGB}{46,117,182}
\definecolor{rmMap}{RGB}{112,173,71}
\usepackage{array}
\usepackage{booktabs}
\usepackage{makecell} 
\usepackage{hyperref}
\usepackage{refcount}

\journal{Nuclear Physics B}

\begin{document}

\begin{frontmatter}



\title{RadMaps: A Geospatial Framework for Simultaneously Modelling Capacity and Geographic Constraints on Radiotherapy Access} 


\author[exceleray]{Laurence Wroe} 
\ead{laurencewroe@gmail.com}

\author[oxford]{Archie Brown}
\author[oxford]{Sophia Martin}
\author[oxford]{Alika Ho}

\author[exceleray,oslo]{Steinar Stapnes} 

\affiliation[exceleray]{
            organization={Exceleray Sàrl},
            city={Geneva},
            postcode={1290}, 
            country={Switzerland}}
\affiliation[oxford]{organization={Department of Physics, University of Oxford},
            addressline={Keble Road}, 
            city={Oxford},
            postcode={OX1 3RH}, 
            country={United Kingdom}}
\affiliation[oslo]{organization={Department of Physics, University of Oslo},
            addressline={Sem Sælands vei 24}, 
            city={Oslo},
            postcode={0371}, 
            country={Norway}}

\begin{abstract}
    \textbf{Background:} Access to radiotherapy is constrained by two compounding factors: insufficient machine capacity to meet patient demand and geographic distance from treatment facilities. Existing analyses address these factors separately, in turn constraining the insights available to advocates, planners and policymakers. This paper presents RadMaps, an open-source\footnote{\label{fn:github}\url{https://github.com/LaurenceWroe/Geospatial-modelling-radiotherapy-access}} geospatial framework that simultaneously models capacity and geographic constraints on radiotherapy access at any spatial scale.\\
    \textbf{Methods:} RadMaps operates on Uber's H3 hexagonal grid and integrates population density data with GLOBOCAN 2024 annual cancer incidence estimates and the IAEA DIRAC facility inventory. Radiotherapy demand is estimated using cancer site specific radiotherapy utilisation rates, and geographic access is modelled via configurable decay functions (step, exponential, Gaussian, or power law) of distance or driving time, with driving times obtained from the TravelTime API. A capped three-step floating catchment area (3SFCA) method allocates demand to facilities subject to both capacity and geographic constraints, producing a localised access metric for every H3 hexagon.\\
    \textbf{Results:} Globally, RadMaps estimates that installed capacity could meet \qty{68.5}{\%} of radiotherapy demand, and \qty{73.0}{\%} of demand is within geographic reach of a facility (\qty{2}{h} driving time with Gaussian decay, H3 Resolution-5), both comparable with published estimates. When both constraints apply, only \qty{60.0}{\%} is met, leaving 3.7~million patients per year untreated. Based on such analysis, countries can be grouped into one of four access profiles: strong access (e.g. the United States), capacity constrained (Japan), geography constrained (Mexico), and doubly constrained (India).\\   
    \textbf{Conclusions:} RadMaps provides a flexible, open-source framework for analysing, visualising and identifying radiotherapy access gaps at regional to global scales, with applications in advocacy, infrastructure planning, and policy prioritisation. RadMaps' modular framework is also readily extensible to other spatial access modelling applications.
\end{abstract}





\end{frontmatter}



\section{Introduction}\label{sec:Introduction}

Cancer is a leading cause of mortality worldwide, with an estimated 20.6 million new cases and 9.8 million deaths in 2024~\citep{Sung_2026}, projected to rise to approximately 35~million new cases by 2050~\citep{Bray_2024}. Radiotherapy, one of the principal treatments for cancer, is indicated for approximately half of all patients for curative, adjuvant, or palliative purposes~\citep{Delaney_2005,Barton_2014}. There is, however, a well-recognised global shortfall in access to radiotherapy, driven by two compounding constraints. 

The first is a \textit{capacity constraint}. Most radiotherapy is delivered as external beam radiotherapy (EBRT), in which a megavoltage machine (MVM), predominantly a linear accelerator (linac), delivers a targeted beam of ionising radiation to the tumour~\citep{Healy_2017}. These machines are resource-intensive: linacs are expensive, require stable local infrastructure, dedicated shielding and ongoing maintenance, and their clinical operation requires a multidisciplinary team of oncologists, medical physicists, dosimetrists, radiation therapists, nurses, and engineers~\citep{Dunscombe_2014}. At a commonly accepted benchmark of 450 patients per machine per year~\citep{Zubizarreta_2015}, the global fleet of approximately 17,000 machines falls well short of demand. Considering capacity alone, we estimate a current shortfall of at least 6,500 machines\footnote{Summing the capacity-only deficit over the 123 countries whose demand exceeds their capacity gives 2.95~million patients per year, equivalent to 6,554 machines at 450 patients per machine per year. This is a lower bound as geographic access is ignored.}, and a recent capacity analysis projected that 16,365 new linacs, together with the replacement of a further 14,105, will be needed by 2045 to keep pace with growing demand~\citep{Moraes_2025}.

The second is a \textit{geographic constraint}. Radiotherapy is typically delivered in daily fractions over several weeks, which places a substantial burden on patients living far from a facility, who may receive suboptimal treatment or none at all~\citep{Silverwood_2024}. A recent geospatial analysis found that \qty{76}{\%} of the global population lives within \qty{2}{h} of travel time to a radiotherapy facility, falling to \qty{17}{\%} in low-income countries~\citep{Wawrzuta_2025}.

Closing these gaps will require substantial investment, with meeting global demand by 2045 estimated to cost US\$124~billion in radiotherapy machines alone~\citep{Moraes_2025}. Cheaper linacs that are more robust in challenging environments are being developed~\citep{Pistenmaa_2018,Dosanjh_2019}, but technological advances alone are not sufficient. Despite its central role in cancer care, radiotherapy accounts for only \qtyrange{4}{12}{\%} of national cancer care budgets in Europe~\citep{Lievens_2020}, and expanding access depends as much on funding, workforce, and political will. Securing them benefits from analyses of radiotherapy access that demonstrate not only how many radiotherapy machines and facilities a country lacks, but where it lacks them, and where additional capacity is most urgently needed.

Existing analyses, however, typically address capacity or geographic constraints alone, although in practice they interact. Capacity-based analyses compare national demand with total machine counts and give a valuable picture of the theoretical shortfall, but do not capture the reduced access of populations far from any facility. Geographic analyses describe the proximity and distribution of facilities but do not account for the finite capacity of each machine. Floating catchment area (FCA) methods combine supply, demand and distance~\citep{Luo_2003,Wan_2012}, and have recently been applied to radiotherapy access within a single region~\citep{Zerrudo_2026}. To our knowledge, however, no existing tool models both constraints simultaneously on a continuous, multi-scale and multi-national spatial framework.

This paper therefore presents RadMaps, an open-source geospatial tool that simultaneously models capacity and geographic constraints on radiotherapy access at any spatial scale, from sub-national to global. RadMaps operates on a hexagonal spatial grid and combines high-resolution population data, national cancer incidence estimates by cancer site, radiotherapy facility inventories, and configurable geographic access decay functions. In the following sections, we describe RadMaps' methodology, analyse its application at global and national scales, compare the results with published capacity-only and geographic-only analyses, identify the constraints that limit access in different countries, and discuss limitations and directions for future work.

\section{Methods}

Ideally, radiotherapy access would be modelled in four steps. First, the rate of new cancer cases would be known by site, stage and patient characteristics at a fine spatial scale throughout the country or region of interest. Second, these cases would be converted into local demand for radiotherapy fractions using the current best-practice, evidence-based optimal radiotherapy utilisation (RTU) rates and treatment plans. Third, the location of every radiotherapy facility would be known, together with its full equipment inventory and the techniques and throughput of each machine. Finally, local access would be calculated by comparing local demand with the capacity of the facilities patients can reach, taking into account how far they are able or willing to travel.

In practice, such data are incomplete or unavailable for most countries. To enable rapid and consistent comparison between regions, RadMaps therefore approximates each step using globally available datasets: national cancer incidence is distributed in proportion to population, assuming uniform incidence per person within each country; demand is calculated using an optimal RTU for each cancer site; demand and capacity are expressed in patients rather than fractions; facilities are taken from a global inventory, with machines assigned a common capacity; and travel burden is represented by a distance-decay function. These approximations can be refined in two ways: users can configure the model inputs --- including the RTUs, machine capacity, and facilities --- for more representative modelling; and, because RadMaps is open-source and modular, its data and methods can be updated and improved collectively.

RadMaps is written in \textsc{Python} and its code is openly available on GitHub\hyperref[fn:github]{\footnotemark[\getrefnumber{fn:github}]}. It is also available as an interactive web application\footnote{\url{https://rtaccess.streamlit.app/}}, built with Streamlit and pydeck on CARTO basemaps~\citep{CartoDB_basemaps}, in which users can select a country or region, configure the model inputs, and visualise and analyse local radiotherapy demand, access ratio, and deficit at their chosen spatial resolution. The modelling pipeline is summarised in \autoref{fig:FlowChart}, and each stage is described in the following subsections.

\begin{figure}[tbh!]
    \centering
    \resizebox{\linewidth}{!}{
\providecolor{rmOutput}{RGB}{237,125,49}
\begin{tikzpicture}[
    font=\fontsize{9}{10.5}\selectfont,   
    box/.style     ={draw, line width=0.9pt, text width=2.7cm, minimum height=1.8cm,   
                     align=center, inner sep=3pt,
                     execute at begin node={\hyphenpenalty=10000\exhyphenpenalty=10000\relax}},
    data/.style    ={box, draw=rmData,   fill=rmData!7},
    input/.style   ={box, draw=rmInput,  fill=rmInput!7},
    metric/.style  ={box, draw=rmMap,    fill=rmMap!9},
    output/.style  ={box, draw=rmOutput, fill=rmOutput!8},
    optional/.style={dashed},
    rowlabel/.style={rotate=90, font=\fontsize{9}{10.5}\selectfont\bfseries, anchor=center, align=center},
    arr/.style     ={-{Latex[length=2mm]}, line width=0.8pt},
    oarr/.style    ={arr, dashed},
  ]
  \def\colsep{3.4}
  \def\yD{0}        
  \def\yU{-2.45}     
  \def\yM{-4.9}     
  \def\yO{-7.1}     
  \def\xL{-1.95}    

  \node[rowlabel, text=rmData]   at (\xL, \yD) {Datasets};
  \node[rowlabel, text=rmInput]  at (\xL, \yU) {Configurable\\ inputs};
  \node[rowlabel, text=rmMap]    at (\xL, \yM) {Per H3\\ hexagon};
  \node[rowlabel, text=rmOutput] at (\xL, \yO) {Outputs};

  \node[data] (kontur) at (0*\colsep, \yD) {Population\\ {\itshape Kontur 2023}};
  \node[data] (gloc)   at (1*\colsep, \yD) {Cancer incidence by site, $I_i$\\ {\itshape GLOBOCAN 2024}};
  \node[data, optional] (rtudat) at (2*\colsep, \yD) {Optimal RTU rates\\ {\itshape CCORE model}};
  \node[data, optional] (tt)     at (3*\colsep, \yD) {Driving times\\ {\itshape TravelTime API}};
  \node[data] (dirac)  at (4*\colsep, \yD) {RT facilities and inventories\\ {\itshape DIRAC 2026}};

  \node[input] (sel)   at (0*\colsep, \yU) {Country / region and H3 resolution};
  \node[input] (rtu)   at (2*\colsep, \yU) {RTU rate, $u_i$\\ (optimal / custom)};
  \node[input] (decay) at (3*\colsep, \yU) {Access variable, $x$, and decay model, $P(x)$};
  \node[input] (cap)   at (4*\colsep, \yU) {Linac capacity, $C_{\text{linac}}$, and optional added facilities};

  \node[metric] (mpop)  at (0*\colsep, \yM) {Population};
  \node[metric] (minc)  at (1*\colsep, \yM) {Annual new cancer cases};
  \node[metric] (mdem)  at (2*\colsep, \yM) {Radiotherapy demand, $D_h$};
  \node[metric] (mgeo)  at (3*\colsep, \yM) {Geographic access probability, $P(x_{hf})$, $P_h$};
  \node[metric] (malloc) at (4*\colsep, \yM) {Allocation (capped 3SFCA): access $A_h$, deficit $\Delta_h$};

  \node[output, text width=9.5cm, minimum height=1.2cm] (out) at (3*\colsep, \yO)
    {Access ratio and deficit maps; national metrics $A$, $A_\text{C}$, $A_\text{G}$, $\rho$ and $\Delta$};

  \draw[arr]  (kontur) -- (sel);
  \draw[arr]  (gloc)   -- (minc);
  \draw[oarr] (rtudat) -- (rtu);
  \draw[oarr] (tt)     -- (decay);
  \draw[arr]  (dirac)  -- (cap);
  \draw[arr]  (cap)    -- (decay);                     

  \draw[arr] (sel)   -- (mpop);                       
  \draw[arr] (rtu)   -- (mdem);
  \draw[arr] (decay) -- (mgeo);
  \draw[arr] (cap)   -- (malloc);

  \draw[arr] (mpop) -- (minc);
  \draw[arr] (minc) -- (mdem);
  \draw[arr] (mdem) -- (mgeo);
  \draw[arr] (mgeo) -- (malloc);

  \draw[arr] (mdem)   -- (mdem   |- out.north);
  \draw[arr] (mgeo)   -- (mgeo   |- out.north);
  \draw[arr] (malloc) -- (malloc |- out.north);

  \draw[line width=0.9pt, dashed] (0.5*\colsep-1.1, \yO-0.15) rectangle ++(0.5, 0.3);
  \node[anchor=west, font=\fontsize{8}{9.5}\selectfont] at (0.5*\colsep-0.5, \yO) {optional dataset / step};
\end{tikzpicture}}
    \caption{RadMaps' modelling pipeline. Datasets (blue) are filtered or parameterised by configurable inputs (purple) to give local, per-hexagon quantities (green), which are combined into access and deficit maps and national metrics (orange). Dashed boxes and arrows denote optional datasets.}
    \label{fig:FlowChart}
\end{figure}

\subsection{Spatial Framework}

RadMaps operates on H3, a hierarchical geospatial indexing system developed by Uber that partitions the globe into hexagonal cells at 16 discrete resolutions~\citep{H3_2025}. For spatial accessibility analysis, H3 offers several advantages over administrative boundaries or rectangular grids: each hexagonal cell has an approximately equal distance from its centroid to all six neighbouring centroids, reducing directional bias; cell areas vary by only about a factor of two across the globe at any resolution; and each cell approximately subdivides into seven children at the next resolution, allowing data to be aggregated efficiently between resolutions. 

RadMaps supports resolutions from H3 Resolution-3 (average cell area of \qty{12400}{km^2}, centroid spacing \qty{120}{km}) to Resolution-6 (\qty{36}{km^2}; \qty{6.5}{km}). 

\subsection{Datasets}~\label{sec:Data}

RadMaps combines up to five datasets, with additional details of their quality and processing given in \ref{app:Data}.
\begin{itemize}
    \item \textbf{Population.} Population estimates for every H3 Resolution-8 cell worldwide are taken from the 2023 Kontur Population Dataset~\citep{kontur_2023} and summed into their parent hexagons at the chosen resolution. 
    \item \textbf{Cancer incidence.} National estimates of annual new cases are taken from GLOBOCAN 2024~\citep{Sung_2026} for 186 countries and territories. RadMaps uses the 33 cancer sites individually reported in GLOBOCAN as well as an `other cancers' category derived from the GLOBOCAN totals (non-melanoma skin cancer is excluded).
    \item \textbf{Radiotherapy facilities.} Radiotherapy facility data are taken from an October 2026 export of the IAEA DIRAC database~\citep{iaea_dirac}, geocoded and processed as described in Appendix~\ref{app:DIRAC}. The processed dataset contains 8,625 radiotherapy centres, 8,501 of which operate at least one MVM, for a total of 17,226 MVMs. Of these, 16,997 lie in the 186 countries and territories with GLOBOCAN estimates and are included in the analysis.\footnote{The remaining 229 are in territories without GLOBOCAN estimates: Taiwan (156), Hong Kong (63), Macao (6), Monaco (2), Aruba (1), and Bermuda (1).} For modelling purposes, all MVMs are assumed to be linacs. The 130 proton and ion therapy systems, 3,301 brachytherapy units, and 707 orthovoltage units recorded in DIRAC are not modelled.
    \item \textbf{Travel time.} Driving times between H3 hexagon centroids and facility locations are obtained from the TravelTime H3 API~\citep{traveltime_2026} under typical traffic conditions departing at 08:00 local time on a Wednesday. They are fetched at H3 Resolution-6 for every hexagon--facility pair within \qty{2}{h} and aggregated to coarser resolutions by population-weighted averaging. For the four countries TravelTime does not cover (China, Russia, North Korea and South Korea), a calibrated straight-line distance is used instead (\autoref{tab:TimeDistance}).
    \item \textbf{Optimal RTUs.} Site-specific optimal RTUs are taken by default from the 2012 update of the Collaboration for Cancer Outcomes Research and Evaluation (CCORE) model~\citep{Barton_2014,Wong_2016}, which estimates that \qty{48.3}{\%} of newly diagnosed cancer patients should receive external-beam radiotherapy (with curative or palliative intent) at least once during their illness. These rates combine guideline-recommended indications for radiotherapy with the proportion of Australian patients meeting each indication. The mapping from the 27 CCORE cancer sites to the 34 RadMaps sites is described in Appendix~\ref{app:RTU}.
\end{itemize}

\subsection{Configurable Inputs}~\label{sec:UserInputs}

RadMaps provides inputs that users can configure for better representation or to explore planning scenarios:
\begin{itemize}
    \item \textbf{Region and resolution.} Users select a country, a region of countries, or the whole world, and an H3 resolution between 3 and 6.

    \item \textbf{Access variable.} Geographic separation $x$ between an H3 hexagon and a radiotherapy facility is measured as either the great-circle distance or the driving time by private car.

    \item \textbf{Distance-decay function.} Longer travel is generally associated with lower treatment uptake and poorer outcomes~\citep{Turner_2023,Baade_2011,Goyal_2014}, although the relationship is complex and context-dependent~\citep{Silverwood_2024,Turner_2026}. RadMaps therefore provides a choice of distance-decay functions $P(x)$:
    \begin{itemize}
        \item Step function: $P(x) = 1$ if $x \leq x_{\max}$, else $P(x) = 0$, where $x_{\max}$ is a hard access limit. It suits analyses against a maximum acceptable travel time~\citep{Luo_2003}.
        \item Exponential: $P(x) = \exp\left(-x / \lambda\right)$, where $\lambda$ is the distance or travel time decay factor at which $P(x)$ falls to $1/e$. It suits analyses where uptake declines steadily with travel~\citep{Wang_2012}.
        \item Gaussian: $P(x) = \bigl(e^{-\frac{1}{2}(x/x_{\max})^2} - e^{-1/2}\bigr)/\bigl(1 - e^{-1/2}\bigr)$ if $x \leq x_{\max}$, else $P(x) = 0$. It suits settings where short journeys barely deter patients but a maximum travel limit applies~\citep{Luo_2009,Dai_2010}.
        \item Power law: $P(x) = \min\bigl(1, (x/x_0)^{-\beta}\bigr)$, where $P(x) = 1$ within $x_0$ and $\beta$ sets the rate of decline beyond it. It can suit settings where long journeys remain common~\citep{Hansen_1959,Joseph_1982}.
    \end{itemize}
    For the exponential and power-law functions, an optional cut-off $x_{\max}$ sets $P(x) = 0$ beyond a maximum travel time or distance.
    
    \item \textbf{Linac capacity.} The number of patients a linac can treat each year depends on operating hours, fractionation schedules, and uptime. RadMaps represents this as a maximum annual throughput $C_\text{linac}$ per machine, set by default to 450 patients per year in line with international benchmarks~\citep{Zubizarreta_2015}. The capacity of individual facilities can be changed by the user.

    \item \textbf{RTUs.} Users can apply the 2012 CCORE optimal rates or enter their own RTUs for each cancer site.

    \item \textbf{Allocation rule.} Capacity is allocated to demand using either a capped three-step floating catchment area (3SFCA) or nearest-first method (\autoref{sec:Allocation}).
    
    \item \textbf{Additional radiotherapy facilities.} Users can add radiotherapy centres and linacs at specified locations, or let RadMaps suggest locations for new facilities. Suggested facilities are placed one at a time at the hexagon with the highest unmet demand or the greatest geographic isolation, up to a fixed number of facilities or until a target is reached for national access, deficit per hexagon, or distance to the nearest facility.
\end{itemize}

\subsection{Radiotherapy access modelling}

\subsubsection{Demand}

RadMaps estimates the annual radiotherapy demand of each country, defined as the number of patients per year who should receive at least one course of external-beam radiotherapy, as
\begin{equation}\label{eq:Demand}
    D = \sum_i u_i I_i,
\end{equation}
where $I_i$ is the annual incidence of cancer site $i$ and $u_i$ is its RTU. This demand is then apportioned to each hexagon $h$ of the country in proportion to its share of the national population as
\begin{equation}\label{eq:DemandHex}
    D_h = D \, \frac{N_h}{N},
\end{equation}
where $N_h$ is the population of hexagon $h$ and $N$ the total population of the country.

\subsubsection{Allocation}~\label{sec:Allocation}

Each facility $f$ has an annual capacity $C_f = n_{f}\,C_\text{linac}$, where $n_{f}$ is its number of linacs, and each hexagon--facility pair has a geographic access probability $P(x_{hf})$, where $x_{hf}$ is the access variable between hexagon $h$ and facility $f$. RadMaps allocates capacity to demand subject to both capacity and geographic constraints using one of two rules: a capped 3SFCA method or a nearest-first rule.

\textbf{Capped 3SFCA.} Floating catchment area (FCA) methods are widely used to measure spatial access to health services~\citep{Luo_2003,Luo_2009}. In the enhanced two-step (E2SFCA) method, as applied here, the ratio of supply to decay-weighted demand is first calculated for each facility as
\begin{equation}\label{eq:E2SFCA}
    R_f = \frac{C_f}{\sum_k P(x_{kf})\,D_k},
\end{equation}
where the sum runs over all hexagons $k$. The access index of each hexagon is then the decay-weighted sum of these ratios over the facilities it can reach, $S^\text{E2SFCA}_h = \sum_f P(x_{hf})\,R_f$. 

Since the demand of a hexagon is counted at every facility it can reach rather than shared between them, demand is inflated where catchments overlap~\citep{Wan_2012,Paez_2019}. The 3SFCA method corrects this by first splitting the demand of each hexagon between the facilities it can reach, using a selection weight $W_{hf} = P(x_{hf}) / \sum_g P(x_{hg})$~\citep{Wan_2012}. The supply-to-demand ratio of each facility then becomes
\begin{equation}\label{eq:3SFCA}
    R_f = \frac{C_f}{\sum_k W_{kf}\,P(x_{kf})\,D_k},
\end{equation}
and the access index of each hexagon is $S^\text{3SFCA}_h = \sum_f W_{hf}\,P(x_{hf})\,R_f$.

To give a number of patients treated rather than a supply-to-demand ratio, RadMaps uses a capped form of 3SFCA, so that no facility treats more than its capacity. The geographically accessible demand of each hexagon is $P_h D_h$, where
\begin{equation}\label{eq:Ph}
    P_h = 1 - \prod_f \bigl(1 - P(x_{hf})\bigr)
\end{equation}
is the probability that a patient can reach at least one facility. This accessible demand is split between facilities using a selection weight
\begin{equation}\label{eq:Selection}
    G_{hf} = \frac{P(x_{hf})\,C_f}{\sum_g P(x_{hg})\,C_g},
\end{equation}
which is the share of the hexagon's accessible demand that seeks treatment at facility $f$. Weighting by $C_f$ follows a Huff-type model~\citep{Huff_1963,Luo_2014} in which patients are attracted to a facility by both its proximity and its size. Each facility $f$ is therefore allocated a demand $Q_f = \sum_h G_{hf} P_h D_h$, and treats the same fraction, $\min(1, C_f/Q_f)$, of each share it receives, giving the number of patients treated in each hexagon as
\begin{equation}\label{eq:A_FCA}
    A_h = P_h D_h \sum_f G_{hf} \min\left(1, \frac{C_f}{Q_f}\right).
\end{equation}
Demand that an over-subscribed facility cannot treat is not redirected to other facilities with spare capacity, and so capped 3SFCA gives a conservative estimate where catchments overlap. The rule assumes that patients spread across all reachable facilities rather than all attending the nearest, consistent with studies showing that a substantial proportion of patients bypass their nearest radiotherapy facility: \qty{60}{\%} of radiotherapy patients in Iowa, United States~\citep{Ward_2014}, and \qty{21}{\%} of men receiving radical prostate radiotherapy in England~\citep{Aggarwal_2017}.

\textbf{Nearest-first.} Hexagon--facility pairs are grouped into equally spaced `rings' of width $\delta x$, set to one hexagon width for distance or \qty{5}{min} for travel time, such that pair $(h,f)$ is assigned to ring $r = \operatorname{round}(x_{hf}/\delta x)$. Each hexagon starts with unmet demand $D^*_h = D_h$ and each facility with remaining capacity $C^*_f = C_f$. Rings are processed in increasing order of $r$ and, within a ring, facilities are processed in a fixed order, sorted by their coordinates\footnote{Where a given hexagon lies in the same ring of more than one facility, the result can depend on the order in which those facilities are processed. Across ten random facility orders, national access under nearest-first allocation changes by 0.5 points on average across the 151 countries with a radiotherapy centre, and by at most 7.6 points (Latvia), while global access changes by 0.1 points.}. For the set of hexagons $H_{rf}$ assigned to ring $r$ of facility $f$, the facility treats a fraction $P(x_{hf})$ of the unmet demand in each, up to its remaining capacity, as
\begin{equation}\label{eq:NF_alloc}
    a_{hf} = P(x_{hf})\,D^*_h \min\left(1, \frac{C^*_f}{\sum_{k \in H_{rf}} P(x_{kf})\,D^*_k}\right), \qquad h \in H_{rf}.
\end{equation}
If the hexagons in a ring exceed the remaining capacity, the remaining capacity is shared in proportion to their decay-weighted unmet demand. The unmet demand and remaining capacity are then updated, $D_h^* \to D_h^* - a_{hf}$ and $C_f^* \to C_f^* - \sum_{h \in H_{rf}} a_{hf}$, before the next ring is processed. 

Once all rings have been processed, the number of patients treated in each hexagon is
\begin{equation}\label{eq:A_NF}
    A_h = \sum_f a_{hf} = D_h - D^*_h.
\end{equation}
This rule assumes that patients attend their nearest facility with available capacity.

\subsubsection{Access metrics}~\label{sec:Limits}

From the number of patients treated in each hexagon, $A_h$, RadMaps derives the local deficit $\Delta_h = D_h - A_h$ and access ratio $\rho_h = A_h/D_h$. Summing over the hexagons of a country, or of a region of several countries, gives
\begin{align}\label{eq:Access_RT}
    A = \sum_h A_h, && \Delta = D - A, && \rho = \frac{A}{D} = \sum_h \frac{D_h}{D}\,\rho_h.
\end{align}

RadMaps is configured such that geography-only and capacity-only metrics can be recovered as:
\begin{itemize}
    \item \textbf{Geography only} ($C_\text{linac} \to \infty$). Every hexagon is treated up to its geographically accessible demand, giving
    \begin{equation}\label{eq:Access_G}
        A_\text{G} = \sum_h P_h D_h.
    \end{equation}
    \item \textbf{Capacity only} ($P(x) \equiv 1$). Every facility can reach all demand in the region, giving
    \begin{equation}\label{eq:Access_C}
        A_\text{C} = \min\Bigl(\sum_f C_f,\; D\Bigr),
    \end{equation}
    which is calculated for each country and summed for regions. 
\end{itemize}

\section{Results} \label{sec:Results}

All results in this section are computed at H3 Resolution-5 using: capped 3SFCA allocation, the 2012 CCORE optimal RTUs, a machine capacity of $C_\text{linac} = 450$ patients per linac per year, and a Gaussian decay with a \qty{2}{h} driving-time limit.\footnote{Driving-time data are not available for China, Russia, North Korea and South Korea. For these countries, a straight-line distance of \qty{64.4}{km}, calibrated to give the same population coverage as a \qty{2}{h} drive, is used instead (Appendix~\ref{app:Travel}).}

\subsection{Global analysis}~\label{sec:Global}

\begin{figure}[tbh!]
  \centering
  \begin{subfigure}{\linewidth}
    \centering
    \includegraphics[width=\linewidth]{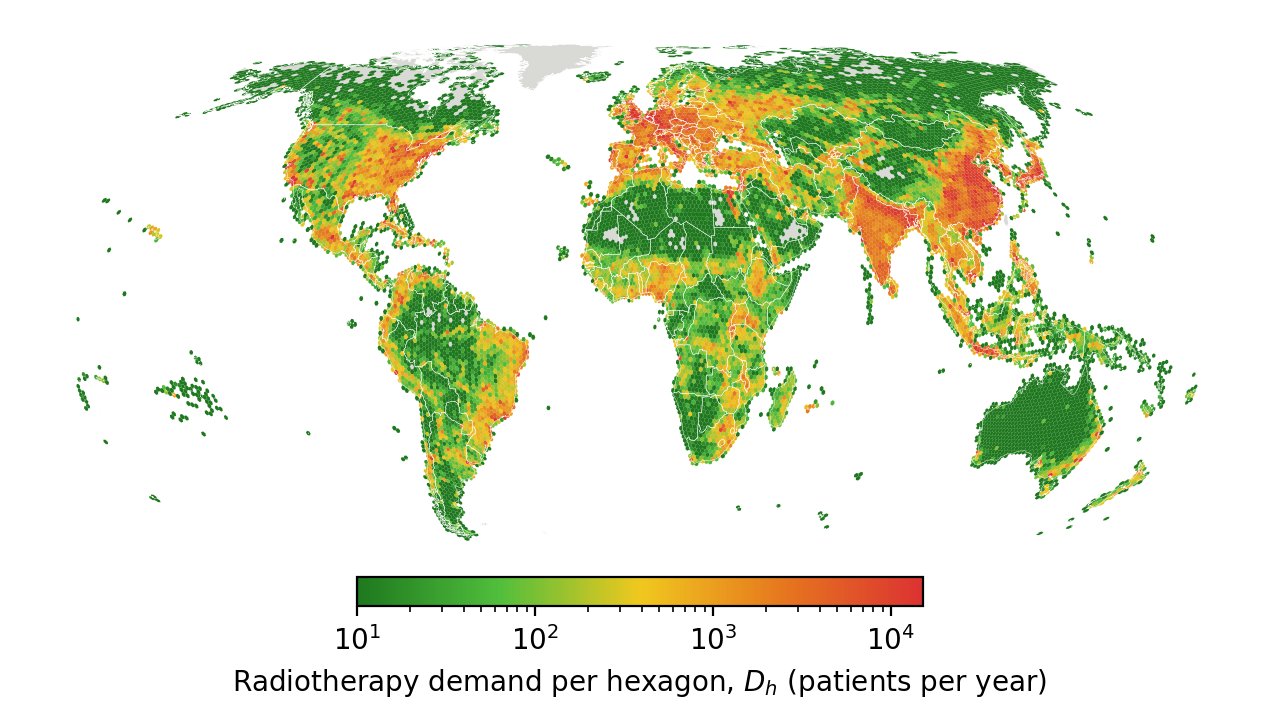}
    \caption{Radiotherapy demand, $D_h$.}
    \label{fig:World_Demand}
  \end{subfigure}
  \begin{subfigure}{\linewidth}
    \centering
    \includegraphics[width=\linewidth]{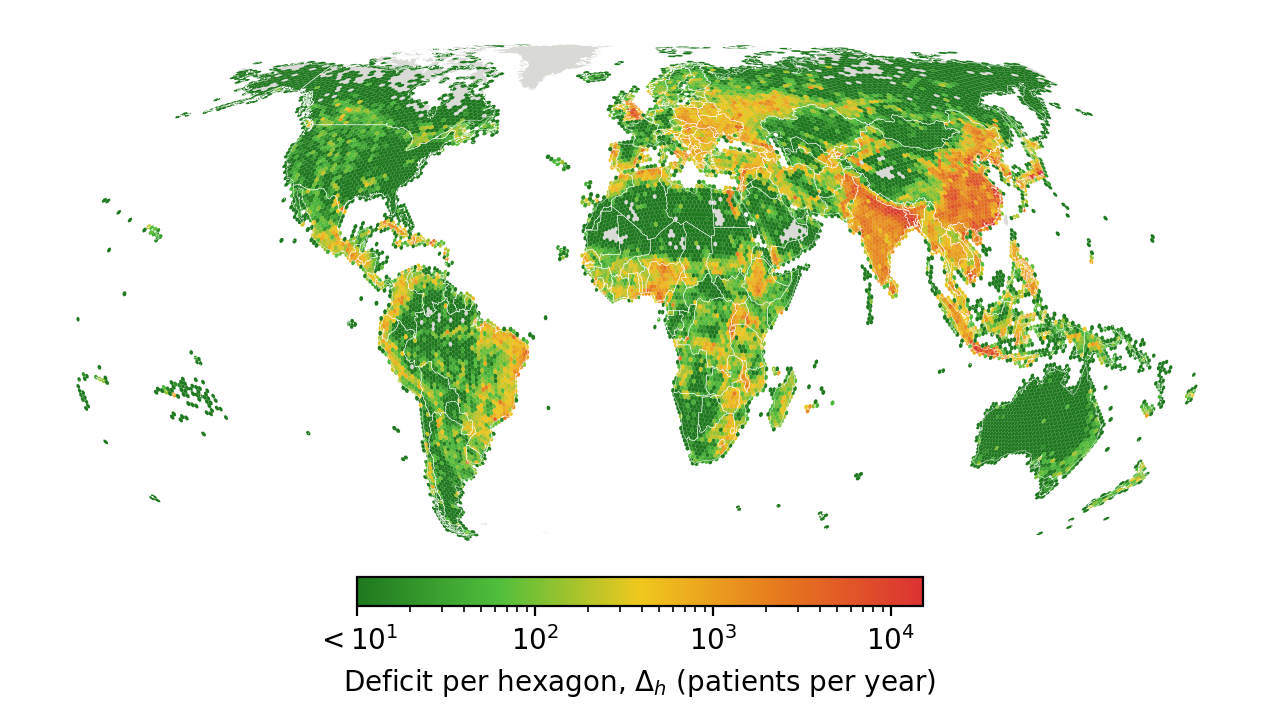}
    \caption{Capped 3SFCA deficit, $\Delta_h$.}
    \label{fig:World_Deficit}
  \end{subfigure}
  \caption{Modelled radiotherapy demand and deficit per H3 Resolution-3 hexagon, summed from the Resolution-5 calculation, on the same logarithmic scale. Countries without GLOBOCAN estimates and hexagons without population are shown in grey.}
  \label{fig:World_Demand_Deficit}
\end{figure}

The total modelled global demand for radiotherapy is 9.4~million patients per year, and \autoref{fig:World_Demand} shows its distribution at H3 Resolution-3.\footnote{The demand and access of each H3 Resolution-5 hexagon are summed into its parent hexagon at Resolution-3, including where the parent spans a national border.} Demand is modelled separately for each country using its national incidence by cancer site, so it reflects both population density and national differences in the rate and mix of cancers.  

\begin{figure}[tbh!]
  \centering
  \includegraphics[width=\linewidth]{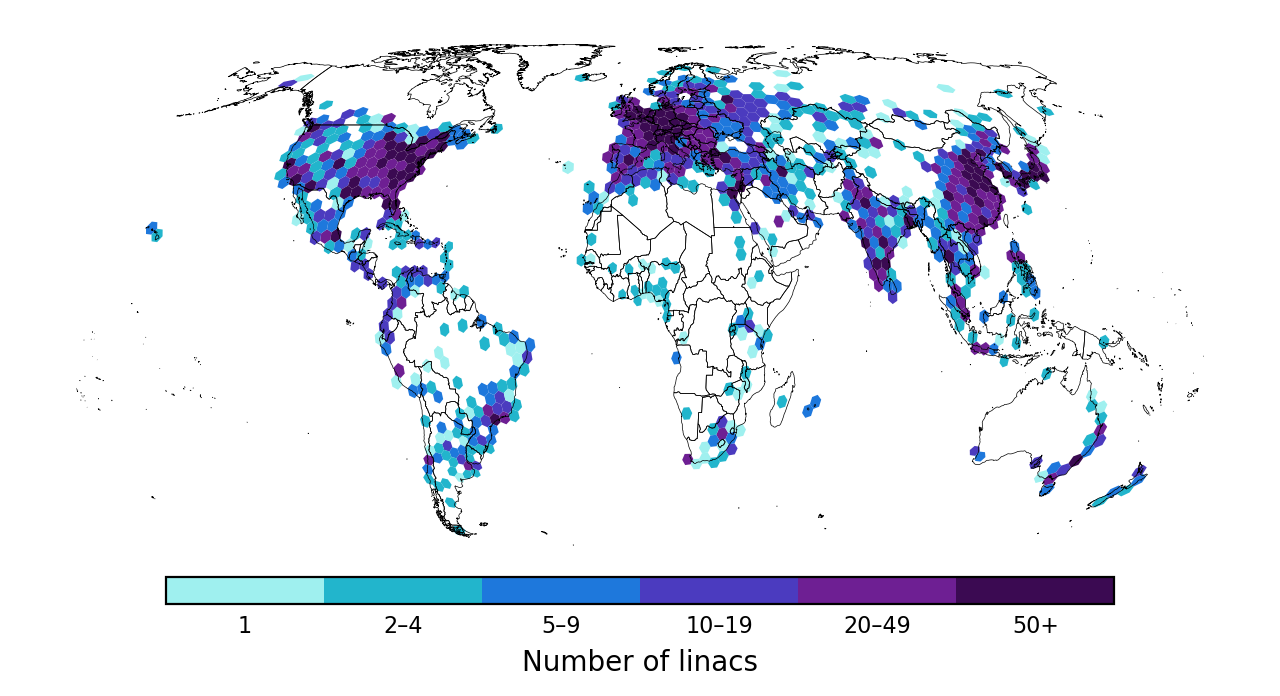}
  \caption{Number of linacs per H3 Resolution-2 hexagon.}
  \label{fig:World_Facilities}
\end{figure}

\autoref{fig:World_Facilities} shows the distribution of the 16,997 linacs at the 8,407 modelled radiotherapy centres at H3 Resolution-2. Thirty-five countries, with a combined demand of 70,000 patients per year (\qty{0.7}{\%} of the total), have no listed centre and are therefore modelled with zero access.

\begin{figure}[tbh!]
  \centering
  \begin{subfigure}{\linewidth}
    \centering
    \includegraphics[width=\linewidth]{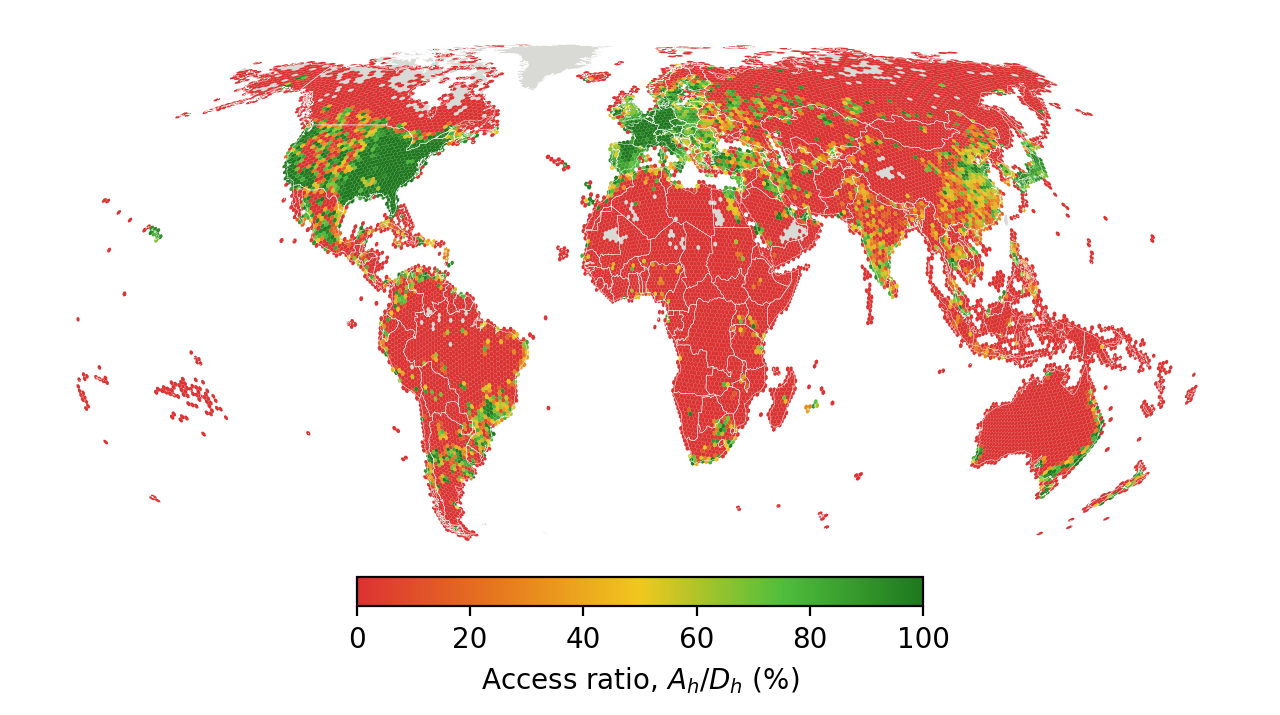}
    \caption{Per H3 Resolution-3 hexagon, $\rho_h = A_h/D_h$, summed from the Resolution-5 calculation.}
    \label{fig:World_Ratio}
  \end{subfigure}
  \begin{subfigure}{\linewidth}
    \centering
    \includegraphics[width=\linewidth]{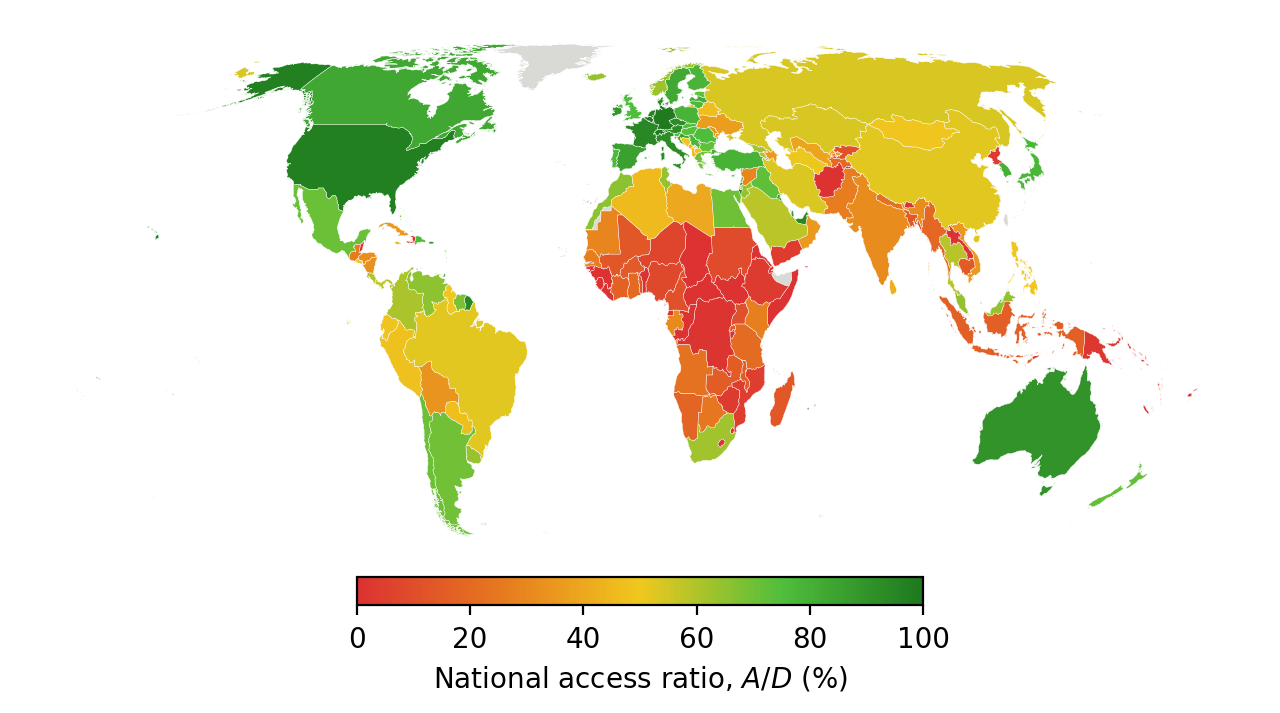}
    \caption{Per country, $\rho = A/D$.}
    \label{fig:World_Choropleth}
  \end{subfigure}
  \caption{Capped 3SFCA access ratio (a) per hexagon and (b) per country, on the same colour scale. Countries without GLOBOCAN estimates, and hexagons without population, are shown in grey.}
  \label{fig:World_Access}
\end{figure}

Combining this distribution of demand and linacs, RadMaps calculates that \qty{60.0}{\%} of demand is met, leaving a total global deficit of 3.7~million patients per year. \autoref{fig:World_Access} shows the calculated access ratio per hexagon and per country. The access ratio $\rho_h$ (\autoref{fig:World_Ratio}) is high around facilities in North America, Europe, eastern China and south-eastern Australia, but low across most of Africa, central and Southeast Asia, and the interiors of South America and Australia. \autoref{fig:World_Choropleth} shows the average national access ratio of every country. 

Many of the large areas of low access, however, are sparsely populated, with too little demand to sustain a radiotherapy centre. The deficit $\Delta_h$ (\autoref{fig:World_Deficit}) scales with demand, and so instead concentrates in densely populated regions with insufficient access. East Asia and the Pacific account for \qty{44}{\%} of the global deficit and South Asia for \qty{18}{\%}, with China and India alone accounting for \qty{45}{\%} (1.1 and 0.6~million patients per year). Sub-Saharan Africa has the lowest access (\qty{15.7}{\%} of demand met), but only \qty{10.5}{\%} of the global deficit because of its relatively lower cancer incidence.

\autoref{tab:Comparison} compares these global RadMaps access metrics with published capacity-only and geography-only metrics, by World Bank income group~\citep{WorldBank_2026}. For capacity only, Moraes et al.~\citep{Moraes_2025} calculate, for 181 countries, the ratio of each country's radiotherapy demand --- derived from GLOBOCAN 2022 incidence and the radiotherapy utilisation rates of Yap et al.~\citep{Yap_2016} --- to the capacity of its linacs listed in DIRAC in 2020--21, assuming 450 patients per machine per year. Applying the same method to our data gives a higher global capacity-only access, \qty{81.6}{\%} against \qty{73.5}{\%}. Since the total demand is similar (9.4 and 9.1~million patients per year), this mostly reflects the larger number of linacs in the DIRAC export used by RadMaps (16,997 compared with 14,792).

For geography only, Wawrzuta et al.~\citep{Wawrzuta_2025} calculate the share of the population within a \qty{2}{h} drive of the nearest domestic radiotherapy centre, routing on OpenStreetMap from the centroids of 178,218 administrative subregions to 8,925 centres verified using a November 2024 DIRAC export. Applying a similar methodology to our 404,480 Resolution-5 hexagons gives agreement to within 6 points for the low-, upper-middle- and high-income groups, but RadMaps is 18.1 points lower for lower-middle-income countries. Almost all of this difference comes from India and Bangladesh, where \qty{60.5}{\%} and \qty{39.2}{\%} of the population lie within a \qty{2}{h} drive in RadMaps, against \qty{87}{\%} and \qty{88}{\%}. This most likely reflects the different routing as TravelTime models typical traffic conditions on congested road networks, whereas OpenStreetMap routing typically assumes free-flowing traffic. The different origins (hexagons rather than subregion centroids) and numbers of centres may also contribute.

\begin{table}[tbh!]
\centering
\caption{Capacity-only, geography-only, and RadMaps calculated access by World Bank income group, as a percentage of demand. $^*$Denotes grouped by the FY2025 World Bank classification, which reproduces the income groups of Ref.~\cite{Moraes_2025}, whereas all other columns use the current FY2027 classification~\cite{WorldBank_2026}. Ref.~\cite{Wawrzuta_2025} values are recomputed from its country-level results with the grouping used here.}
\label{tab:Comparison}
\small
\begin{tabular}{lrrrrrrr}
\toprule
 & \multicolumn{2}{c}{Capacity only$^*$} & \multicolumn{2}{c}{Geography only} & \multicolumn{3}{c}{RadMaps} \\
\cmidrule(lr){2-3}\cmidrule(lr){4-5}\cmidrule(lr){6-8}
\makecell[l]{Income\\group} & \makecell[r]{Ref.~\cite{Moraes_2025}\\(\%)} & \makecell[r]{RadMaps\\(\%)} & \makecell[r]{Ref.~\cite{Wawrzuta_2025}\\(\%)} & \makecell[r]{RadMaps\\(\%)} & \makecell[r]{$A_\text{C}/D$\\(\%)} & \makecell[r]{$A_\text{G}/D$\\(\%)} & \makecell[r]{$A/D$\\(\%)} \\
\midrule
Low & 7.2 & 7.8 & 16.3 & 13.6 & 7.7 & 11.5 & 6.4 \\
Lower-middle & 36.1 & 42.3 & 69.9 & 51.8 & 40.7 & 38.5 & 30.0 \\
Upper-middle & 46.6 & 62.6 & 85.4 & 79.5 & 59.9 & 72.3 & 51.0 \\
High & 121.1 & 123.8 & 92.8 & 93.2 & 93.2 & 92.1 & 85.4 \\
\midrule
World & 73.5 & 81.6 & 75.0 & 66.1 & 68.5 & 73.0 & 60.0 \\
\bottomrule
\end{tabular}
\end{table}

Comparing the RadMaps access metrics in \autoref{tab:Comparison}, the combined access $A$ lies below both single-constraint metrics: globally, \qty{60.0}{\%} of demand is met, 8.5 points below the lower of $A_\text{C}/D$ and $A_\text{G}/D$, leaving a further 0.8~million patients per year untreated. This is because the two constraints compound: capacity is not always located where patients can reach it, and centres within reach of patients may already be at capacity. Modelling capacity and geography simultaneously is therefore necessary, as either single-constraint metric overestimates the number of patients treated.

A sensitivity analysis in \ref{app:Sensitivity} shows that global access, \qty{60.0}{\%} in the main configuration, falls to \qty{28.0}{\%} with a \qty{30}{min} travel-time limit and to \qty{42.3}{\%} with a capacity of 250 patients per linac per year. It ranges from \qty{53.9}{\%} (power law) to \qty{62.4}{\%} (step function) across decay models, is \qty{59.0}{\%} with nearest-first allocation, and varies by at most 0.3 points between H3 Resolution-3 and Resolution-5.

\subsection{Individual country analysis}~\label{sec:Countries}

\begin{figure}[tbh!]
  \centering
  \begin{subfigure}{0.49\linewidth}
    \centering
    \includegraphics[width=\linewidth]{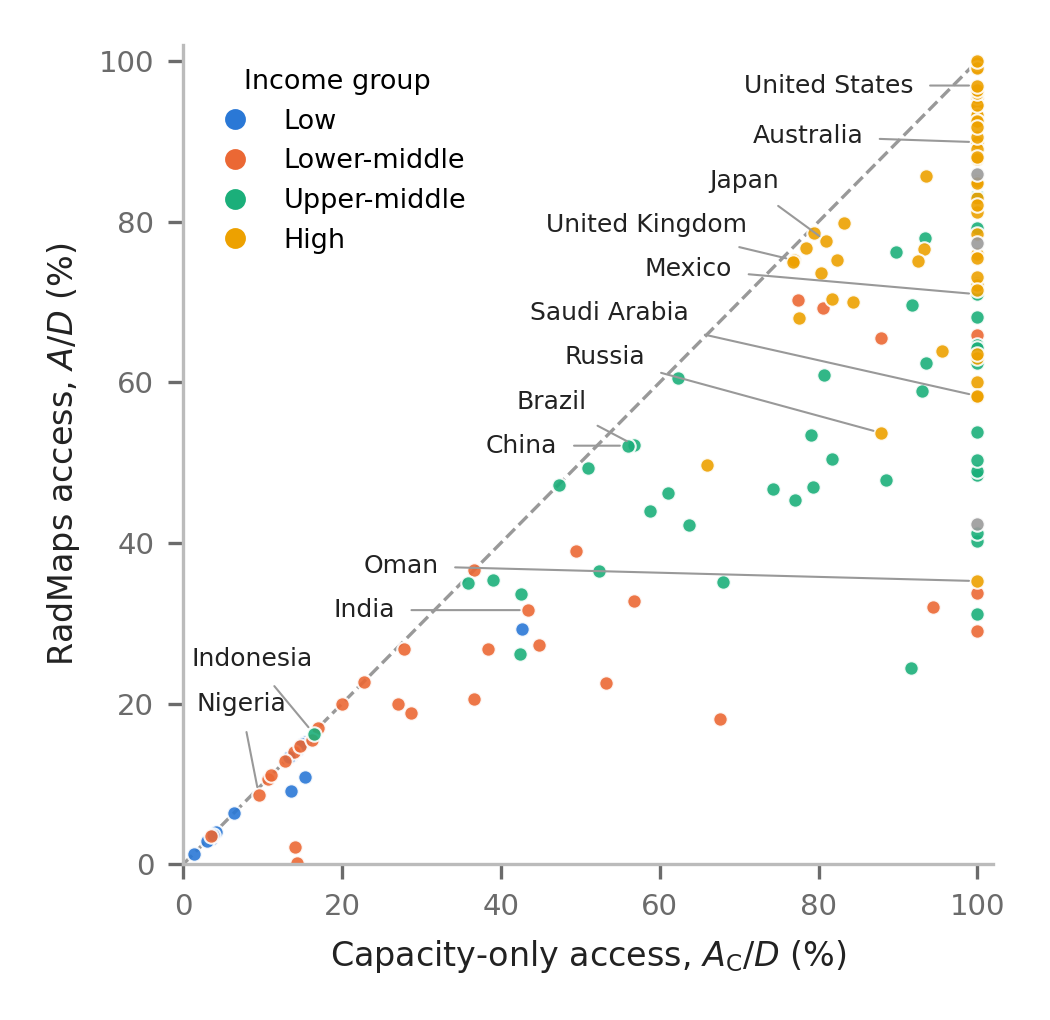}
    \caption{Against capacity-only access, $A_\text{C}/D$.}
    \label{fig:Country_Capacity}
  \end{subfigure}\hfill
  \begin{subfigure}{0.49\linewidth}
    \centering
    \includegraphics[width=\linewidth]{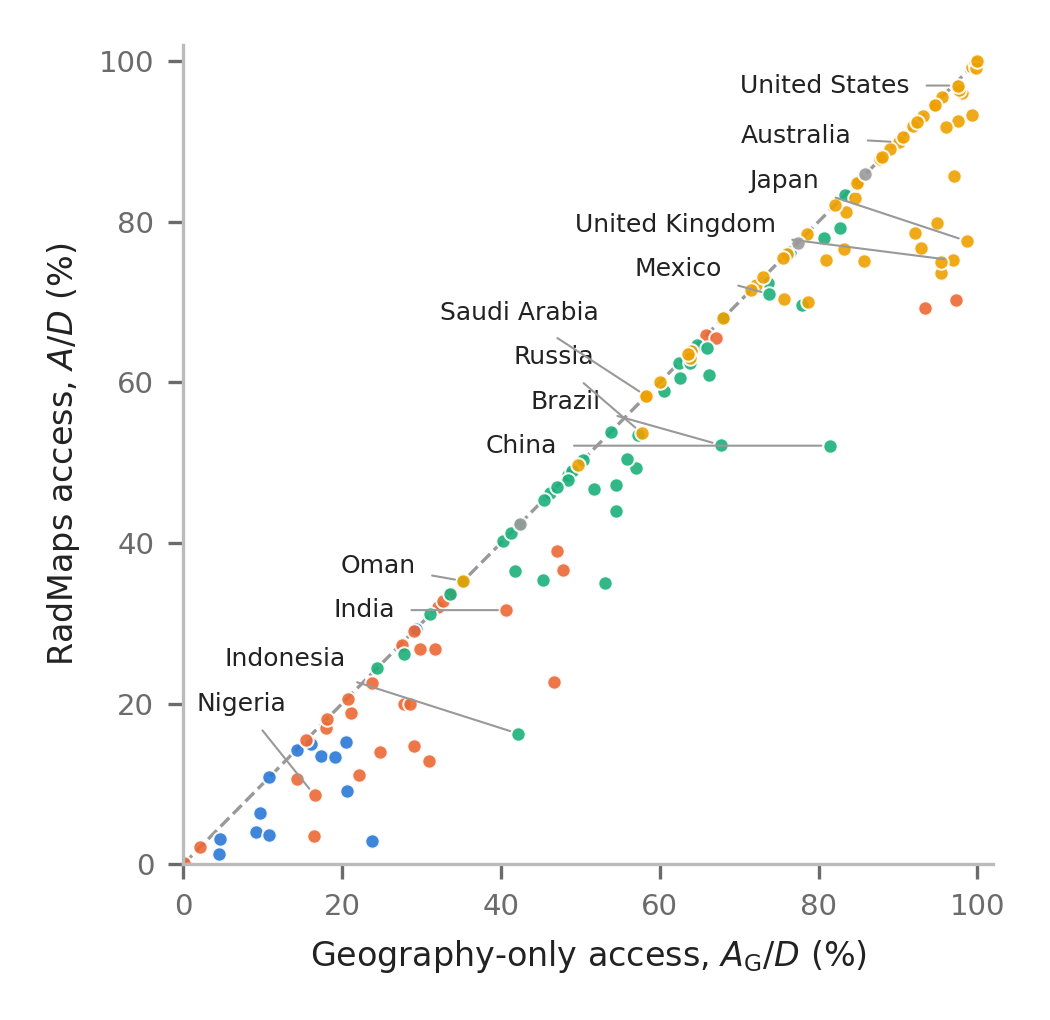}
    \caption{Against geography-only access, $A_\text{G}/D$.}
    \label{fig:Country_Geography}
  \end{subfigure}
  \caption{RadMaps access $A/D$ of each country with at least one linac against (a) capacity-only and (b) geography-only access, coloured by World Bank income group.}
  \label{fig:Country_Comparison}
\end{figure}

By comparing the combined access $A/D$ with the capacity-only $A_\text{C}/D$ and geography-only $A_\text{G}/D$ metrics (\autoref{fig:Country_Comparison}), countries can be grouped into four categories using a nominal threshold of \qty{80}{\%} of demand:
\begin{itemize}
    \item \textbf{Strong access.} ($A/D \geq \qty{80}{\%}$; 31 countries, \qty{24}{\%} of global demand). Countries where most demand is met, such as the United States and Australia.
    \item \textbf{Capacity constrained.} ($A/D < \qty{80}{\%}$, $A_\text{C} < A_\text{G}$; 10 countries, \qty{36}{\%} of global demand). Countries whose radiotherapy centres are well distributed across the population but have insufficient capacity, such as the United Kingdom, China and Japan.
    \item \textbf{Geography constrained.} ($A/D < \qty{80}{\%}$, $A_\text{G} < A_\text{C}$; 51 countries, \qty{13}{\%} of global demand). Countries with sufficient national capacity concentrated in too few locations to reach the population, such as Mexico, Saudi Arabia, Oman and Russia.
    \item \textbf{Doubly constrained.} ($A_\text{C}/D < \qty{80}{\%}$ and $A_\text{G}/D < \qty{80}{\%}$; 94 countries, \qty{28}{\%} of global demand). Countries with neither sufficient capacity nor sufficient geographic coverage, such as India, Brazil, Indonesia and Nigeria, and including the 35 countries with no radiotherapy centre.
\end{itemize}

Each category is illustrated in \ref{app:Maps} by maps of demand, linac distribution, access ratio and deficit for the United States (strong access), Japan (capacity constrained), Mexico (geography constrained), and India (doubly constrained). Such maps support not only the analysis of radiotherapy access, but also infrastructure planning and advocacy.

\section{Discussion}~\label{sec:Discussion}

RadMaps demonstrates the value of simultaneously modelling capacity and geographic constraints on radiotherapy access to produce a combined metric that reveals access deficits not captured by either constraint alone. The global analysis illustrates this clearly: the capacity-only access of \qty{68.5}{\%} and geography-only access of \qty{73.0}{\%} simulated together yield a combined access of only \qty{60.0}{\%}, underscoring how the two constraints compound in ways that neither metric alone can capture. The national analyses further demonstrate the tool's ability to distinguish qualitatively distinct access profiles and localise deficits at sub-national scale, pointing towards whether investment should prioritise additional machines at existing centres, new facility locations, or both.

RadMaps nonetheless relies on a number of assumptions that should be considered when interpreting results.

\textbf{Uniform cancer incidence.} RadMaps distributes cancer incidence uniformly across the population, such that the radiotherapy demand within each hexagon is directly proportional to its share of the national population. In practice, cancer risk varies spatially within countries due to demographic differences (for example, older, more urbanised, or more deprived populations tend to have higher incidence rates~\citep{Brown_2018,Wild_2020}) meaning that demand may be systematically under- or over-estimated in certain regions. This limitation could be addressed by incorporating sub-national cancer incidence data where available.

\textbf{No patient stratification.} All radiotherapy demand is treated as equivalent, whereas in practice the treatment required varies with cancer site, disease stage, patient age, mobility, and due to socioeconomic reasons. These factors alter both the demand for radiotherapy and the access achievable, and stratifying demand by them would improve the model.

\textbf{Dataset limitations.} The global datasets enable rapid and consistent modelling, but may be incomplete, out of date or unrepresentative. For example, GLOBOCAN national cancer incidences for 6 countries and territories with radiotherapy facilities, including Hong Kong and Taiwan, are not available; the 2012 CCORE optimal RTUs are based on guidelines that may have since been superseded; DIRAC is voluntarily reported and may not fully capture recent facility openings or closures~\citep{Atun_2015}; and TravelTime's H3 API does not serve China, Russia, North Korea or South Korea, for which a calibrated straight-line distance is used instead.

\textbf{Linac capacity.} RadMaps uses a machine capacity based on patients rather than fractions per linac per year, by default assuming a throughput of 450 patients per linac per year for every country~\citep{Zubizarreta_2015}. In practice, throughput is determined by the number of fractions delivered, so fixing this number implicitly assumes a case mix and the fractionation schedules prescribed for it. The 450 figure is also derived from guidelines more than fifteen years old~\citep{Slotman_2005,IAEA_2010}, which predate the widespread adoption of stereotactic radiotherapy and moderate and ultra-hypofractionation. Deriving capacity from delivered fractions, using each country's own case mix and current fractionation practice, would make the modelling more representative. 

\textbf{Interpretation.} RadMaps results are modelled estimates of access. Validation against national or regional utilisation data is required to establish the tool's predictive validity.

Additional assumptions worth noting include: only incident cancer cases are modelled, excluding patients requiring re-treatment; all MVMs are treated as equivalent linacs; national borders are treated as hard limits on patient flow, so cross-border treatment is not modelled; private and public facilities are treated equally; travel is represented by driving times for a single weekday-morning departure, although many patients rely on public transport; and, under capped 3SFCA, demand that an over-subscribed facility cannot treat is not redirected to facilities with spare capacity. 

Notwithstanding these limitations, RadMaps provides a flexible and computationally efficient framework for radiotherapy access modelling that can be applied consistently across any country or region. Its modular design allows the underlying datasets and model parameters to be updated as new data are released or more representative data become available, and we intend to keep RadMaps updated with each new release of GLOBOCAN, DIRAC and the Kontur population dataset. Furthermore, the tool is freely accessible via a Streamlit web application and open-source GitHub repository and, beyond radiotherapy, its modelling framework is readily adaptable to access modelling for other healthcare facilities and services where capacity and geographic constraints interact.

\section{Conclusion}

RadMaps is an open-source geospatial framework that simultaneously models capacity and geographic constraints on radiotherapy access, producing a combined access metric that captures the compounding effect of both constraints at sub-national scales. Applied globally, it estimates that only \qty{60.0}{\%} of radiotherapy demand is met, below both the capacity-only (\qty{68.5}{\%}) and geography-only (\qty{73.0}{\%}) estimates. By separating countries limited by capacity, by geography or by both, and showing where within each country demand goes unmet, RadMaps indicates whether additional machines, new facility locations or both are needed, providing evidence to support planning, funding and advocacy for radiotherapy. The modular framework is freely available and readily extensible, providing a foundation for larger-scale analyses of radiotherapy access and for access modelling of other services across diverse geographic and socioeconomic contexts.

While this paper describes the methodology underpinning RadMaps, the tool is being distributed to healthcare professionals and radiotherapy planners to evaluate its practical utility and to validate modelled access against observed treatment uptake. This engagement will inform future development including sub-national incidence data, public transport times, redirection of patients between facilities and projections of future demand, and will establish the evidence base for the tool's use in infrastructure planning and policy prioritisation.

\section*{Declarations}

\subsection*{Acknowledgements}

The authors are thankful to TravelTime who provided extended API-usage and assistance with incorporating their data into the model, and to Walter Wuensch for feedback on the manuscript.

\subsection*{Data Availability}

This study uses publicly accessible data. The RadMaps framework is available on GitHub \url{https://github.com/LaurenceWroe/Geospatial-modelling-radiotherapy-access}.

\subsection*{Ethics approval and consent to participate}
Not applicable.

\subsection*{Consent for publication}
Not applicable.

\subsection*{Competing interests}
The authors declare no competing interests.

\subsection*{Funding}
Not applicable.

\subsection*{Authors' contributions}

LW: Conceptualisation; Data curation; Investigation; Methodology; Supervision; Validation; Visualisation; Writing - original draft; Writing - review \& editing.

AB: Data curation; Investigation; Visualisation; Writing - review \& editing.

SM: Data curation; Investigation; Visualisation; Writing - review \& editing.

AH: Data curation; Investigation; Visualisation; Writing - review \& editing.

SS: Supervision; Writing - review \& editing.


\bibliographystyle{elsarticle-num}
\bibliography{cas-refs}

\clearpage\appendix
\renewcommand{\thesubsection}{\Alph{section}.\arabic{subsection}}

\section{Data Source Details}~\label{app:Data}

This appendix provides additional details behind the five datasets in RadMaps v1.1. 

\subsection{Population}~\label{app:Pop}

The Kontur Population Dataset released on November~2023~\citep{kontur_2023} provides a population estimate for every H3 Resolution-8 cell (mean area \qty{0.74}{km^2}) worldwide, built from the Global Human Settlement Layer, OpenStreetMap and related source estimates between March 2020 and November 2023 and calibrated to national totals. Kontur clips cells to national boundaries, so a cell straddling a border appears once per country with that country's share of its population. The global total population included in the dataset is 8.02~billion. 

\subsection{Cancer incidence (GLOBOCAN~2024)}~\label{app:Cancer}

National new-case estimates for 33 cancer sites\footnote{Bladder; brain and central nervous system; breast; cervix uteri; colorectum; corpus uteri; gallbladder; Hodgkin lymphoma; hypopharynx; Kaposi sarcoma; kidney; larynx; leukaemia; lip and oral cavity; liver; lung; melanoma; mesothelioma; multiple myeloma; nasopharynx; non-Hodgkin lymphoma; oesophagus; oropharynx; ovary; pancreas; penis; prostate; salivary glands; stomach; testis; thyroid; vagina; vulva. Two further categories, non-melanoma skin cancer (all cancers minus all cancers excluding non-melanoma skin cancer) and ``other cancers'' (all cancers excluding non-melanoma skin cancer minus the sum of the listed sites), are derived from the data.} are taken from GLOBOCAN~2024~\citep{Sung_2026}. All 186 countries and territories for which GLOBOCAN publishes estimates are modelled. A further 24 territories in the RadMaps world region have a Kontur population file but no GLOBOCAN estimate\footnote{Andorra, Antigua and Barbuda, Aruba, Bermuda, British Virgin Islands, Dominica, Grenada, Hong Kong, Kiribati, Liechtenstein, Macao, Marshall Islands, Micronesia, Monaco, Nauru, Palau, Saint Kitts and Nevis, Saint Vincent and the Grenadines, San Marino, Taiwan, Tonga, Turks and Caicos Islands, Tuvalu and the Holy See.} and are omitted from RadMaps analysis.

\subsection{Radiotherapy facilities (IAEA DIRAC)}~\label{app:DIRAC}

The IAEA Directory of Radiotherapy Centres export of October~2026 used here contains 8,630 entries. Five of these entries were removed as duplicates, leaving 8,625 centre records in 157 countries, as summarised in \autoref{tab:DIRAC}.

\begin{table}[tbh!]
    \centering
    \caption{Composition of the DIRAC export used by RadMaps.}
    \label{tab:DIRAC}
    \small
    \setlength{\tabcolsep}{6pt}
    \renewcommand{\arraystretch}{1.2}
    \begin{tabular}{@{}lrr@{}}
        \toprule
        \textbf{Records} & \textbf{Centres} & \textbf{MVMs} \\ \midrule
        Total records (157 countries and territories) & 8,625 & 17,226 \\
        \quad with at least one megavoltage photon/electron machine (MVM) & 8,501 & 17,226 \\
        \quad with only brachytherapy, orthovoltage, or particle therapy only & 124 & -- \\ \midrule
        \textbf{Modelled in RadMaps (151 countries)} & \textbf{8,407} & \textbf{16,997} \\ \midrule
        Not modelled: MVMs in territories without GLOBOCAN estimates$^*$ & 94 & 229 \\
        Not modelled: proton or ion therapy  & 124 & -- \\
        Not modelled: orthovoltage  & 629 & -- \\
        Not modelled: brachytherapy units & 2,691 & -- \\ \midrule
        Records with a last-update year of 2023--2026 & \multicolumn{2}{r}{\qty{51}{\%}} \\
        \quad 2020--2022 & \multicolumn{2}{r}{\qty{21}{\%}} \\
        \quad 2016--2019 & \multicolumn{2}{r}{\qty{23}{\%}} \\
        \quad 2015 or earlier & \multicolumn{2}{r}{\qty{5}{\%}} \\
        \bottomrule
    \end{tabular}

    \vspace{2pt}
    {\footnotesize $^*$Taiwan (73 centres, 156 MVMs), Hong Kong (16, 63), Macao (2, 6), Monaco (1, 2), Aruba (1, 1) and Bermuda (1, 1).}
\end{table}

Of the 186 modelled countries, 151 have at least one usable facility. The remaining 35\footnote{Afghanistan, Belize, Benin, Bhutan, Burundi, Cabo Verde, Central African Republic, Chad, Comoros, Congo, Djibouti, Equatorial Guinea, Eritrea, Eswatini, Fiji, French Guiana, French Polynesia, Gambia, Guam, Guinea, Guinea-Bissau, Haiti, Lesotho, Liberia, Maldives, New Caledonia, Saint Lucia, Samoa, Sao Tome and Principe, Sierra Leone, Solomon Islands, Somalia, South Sudan, Timor-Leste and Vanuatu.} have no registered MVM and so are modelled with zero capacity.

The database records the name, city, and country of each centre. To determine their coordinates, records were geocoded using these address fields against OpenStreetMap and Google Maps, and every coordinate was checked to lie within the Natural Earth polygon of its stated country. The geocoded export and the record-level audit log are available on the GitHub. Location uncertainty is at the scale of cities, with the 8,501 modelled facilities occupying 7,730 distinct coordinate locations. Because the capped 3SFCA rule shares the demand of each hexagon between facilities in proportion to their capacity, co-located records with the same travel times are treated exactly as a single merged facility.

\subsection{Travel time (TravelTime API)}~\label{app:Travel}

Driving times from each hexagon centroid to each facility are obtained from the TravelTime H3 API~\citep{traveltime_2026}, which snaps centroids to 2026 road networks and returns the mean travel time per cell. Requests assume departure at 08:00 local time on a Wednesday and travel under typical traffic conditions. Times are requested at H3 Resolution-6 with a cut-off of two hours, and only the hexagon--facility pairs reachable within the cut-off are stored. Cells beyond the cut-off, and cells whose centroid cannot be snapped to a routable road (uninhabited terrain, islands without road access), are recorded as unreachable and receive zero geographic access. Coarser resolutions are derived from the Resolution-6 times by population-weighted averaging over each parent's children. Here, a child that cannot reach a facility is counted at the cut-off and a parent is unreachable only if none of its children is.

Travel-time estimates are not available for four of the modelled countries (China, Russia, North Korea and South Korea). For these, the global analyses substitute a straight-line distance threshold chosen so that, across the 152 countries with driving-time data, the population-weighted geographic access at that distance equals the access at the corresponding driving time (\autoref{tab:TimeDistance}).

\begin{table}[tbh!]
    \centering
    \caption{Straight-line distance equivalent to each driving-time threshold, fitted on the countries with TravelTime coverage.}
    \label{tab:TimeDistance}
    \small
    \setlength{\tabcolsep}{8pt}
    \renewcommand{\arraystretch}{1.2}
    \begin{tabular}{@{}cc@{}}
        \toprule
        \textbf{Driving time (min)} & \textbf{Equivalent distance (km)} \\ \midrule
        30  & 12.6 \\
        60  & 28.4 \\
        90  & 41.6 \\
        120 & 64.4 \\
        \bottomrule
    \end{tabular}
\end{table}

\subsection{Radiotherapy utilisation rates}~\label{app:RTU}

Optimal RTU rates are taken from the 2012 update of the CCORE model~\citep{Barton_2014}, with per-site values as tabulated by Wong \textit{et al.}~\citep{Wong_2016}. These rates count curative and palliative first courses; re-treatment is not included. 

The 27 CCORE sites are mapped to the GLOBOCAN sites by assigning the head-and-neck rate to its six subsites (lip and oral cavity, salivary glands, oropharynx, nasopharynx, hypopharynx and larynx), the lymphoma rate to Hodgkin and non-Hodgkin lymphoma, and the ``other'' rate to Kaposi sarcoma, mesothelioma, penis and the derived ``other cancers'' category. The GLOBOCAN colorectum category receives the colon (\qty{4}{\%}) and rectum (\qty{60}{\%}) rates weighted by their Australian incidence (8.4:4.2), giving \qty{23}{\%}. The CCORE unknown-primary site (\qty{61}{\%}) has no GLOBOCAN equivalent; these cases fall within ``other cancers''. Non-melanoma skin cancer is not notifiable in Australia and is excluded from the CCORE model~\citep{Wong_2016}, so it is assigned no radiotherapy demand. The resulting rates for each GLOBOCAN site are listed in \autoref{tab:RTU_RadMaps}.

\begin{table}[tbh!]
    \centering
    \caption{Optimal radiotherapy utilisation rate (\% of new cases) used by RadMaps for each GLOBOCAN cancer site, from the 2012 CCORE model~\citep{Barton_2014,Wong_2016}. Unmarked sites take the rate of the corresponding CCORE site directly. Marked sites have no direct CCORE counterpart and take the rate of: $^a$head and neck; $^b$lymphoma; $^c$colon and rectum, weighted by Australian incidence (8.4:4.2); $^d$other; $^e$not included in the CCORE model.}
    \label{tab:RTU_RadMaps}
    \small
    \setlength{\tabcolsep}{6pt}
    \renewcommand{\arraystretch}{1.1}
    \begin{tabular}{@{}lr@{\hspace{18pt}}lr@{\hspace{18pt}}lr@{}}
        \toprule
        \textbf{Site} & \textbf{RTU (\%)} & \textbf{Site} & \textbf{RTU (\%)} & \textbf{Site} & \textbf{RTU (\%)} \\ \midrule
        Bladder & 47 & Leukaemia & 4 & Pancreas & 49 \\
        Brain, CNS & 80 & Lip, oral cavity & 74$^a$ & Penis & 19$^d$ \\
        Breast & 87 & Liver & 0 & Prostate & 58 \\
        Cervix uteri & 71 & Lung & 78 & Salivary glands & 74$^a$ \\
        Colorectum & 23$^c$ & Melanoma & 21 & Stomach & 27 \\
        Corpus uteri & 32 & Mesothelioma & 19$^d$ & Testis & 15 \\
        Gallbladder & 17 & Multiple myeloma & 45 & Thyroid & 4 \\
        Hodgkin lymphoma & 73$^b$ & Nasopharynx & 74$^a$ & Vagina & 94 \\
        Hypopharynx & 74$^a$ & NHL & 73$^b$ & Vulva & 39 \\
        Kaposi sarcoma & 19$^d$ & Oesophagus & 71 & Other cancers & 19$^d$ \\
        Kidney & 15 & Oropharynx & 74$^a$ & NMSC & 0$^e$ \\
        Larynx & 74$^a$ & Ovary & 4 &  &  \\
        \bottomrule
    \end{tabular}
\end{table}

\clearpage
\section{Sensitivity Analysis}~\label{app:Sensitivity}

\begin{figure}[tbh!]
  \centering
  \begin{subfigure}{0.49\linewidth}
    \centering
    \includegraphics[width=\linewidth]{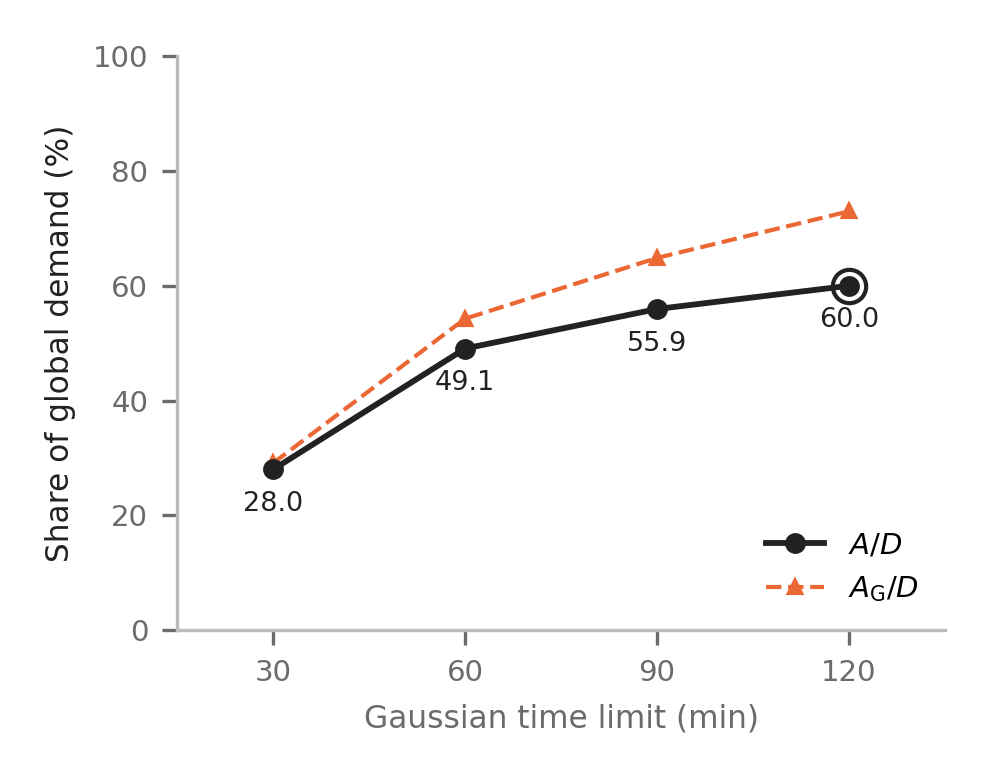}
    \caption{Gaussian travel-time limit.}
    \label{fig:Sensitivity_Time}
  \end{subfigure}
  \begin{subfigure}{0.49\linewidth}
    \centering
    \includegraphics[width=\linewidth]{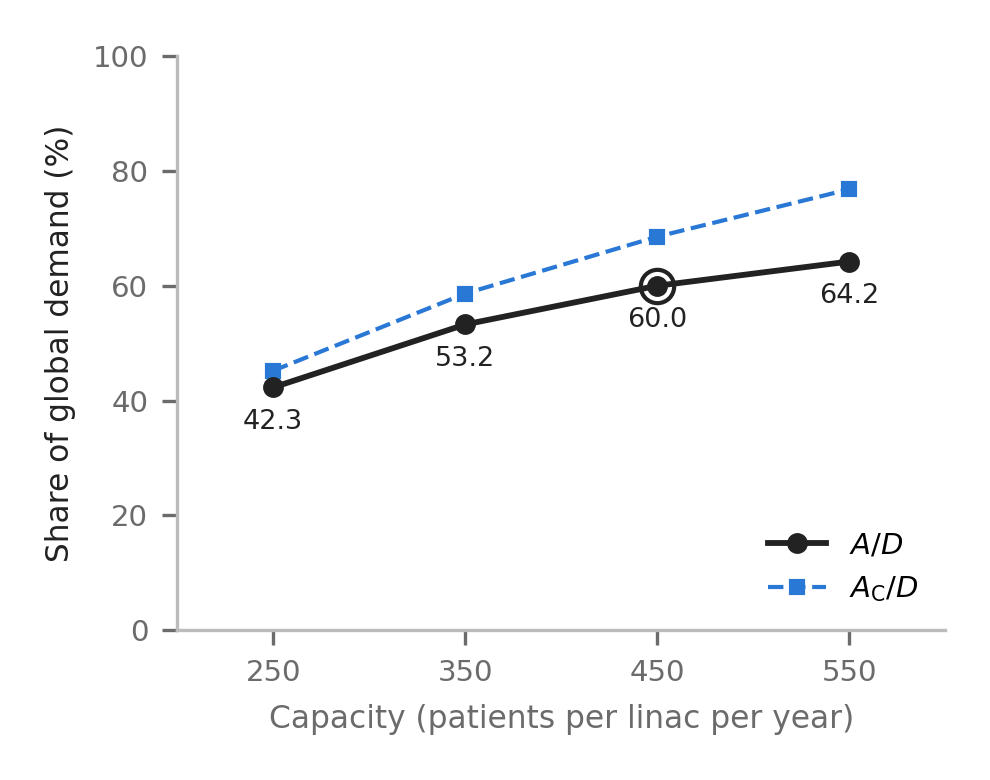}
    \caption{Machine capacity.}
    \label{fig:Sensitivity_Capacity}
  \end{subfigure}\hfill

  \vspace{4pt}
  \begin{subfigure}{0.49\linewidth}
    \centering
    \includegraphics[width=\linewidth]{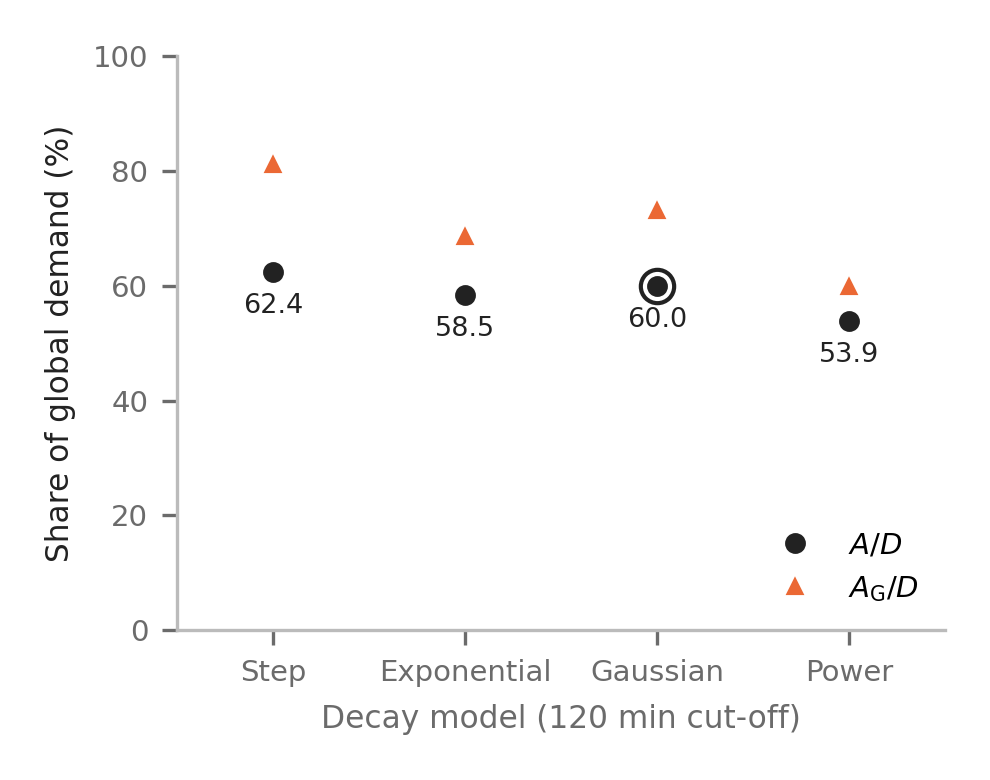}
    \caption{Decay model.}
    \label{fig:Sensitivity_Decay}
  \end{subfigure}
  \begin{subfigure}{0.49\linewidth}
    \centering
    \includegraphics[width=\linewidth]{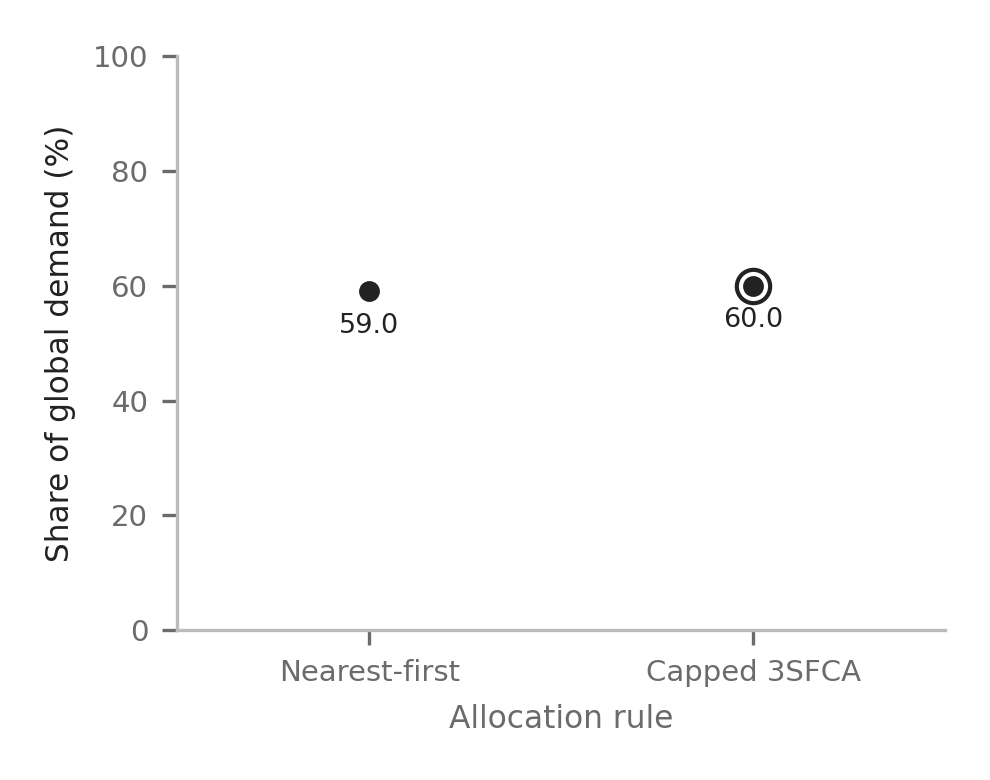}
    \caption{Allocation rule.}
    \label{fig:Sensitivity_Allocation}
  \end{subfigure}\hfill

  \vspace{4pt}
  \begin{subfigure}{0.49\linewidth}
    \centering
    \includegraphics[width=\linewidth]{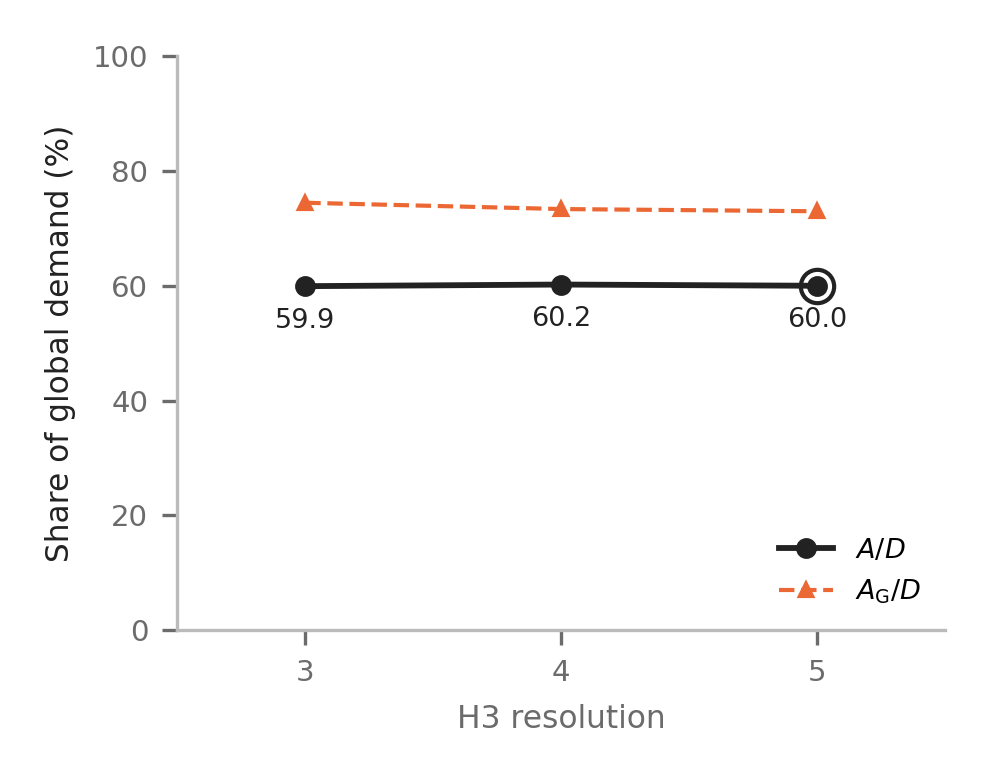}
    \caption{H3 resolution.}
    \label{fig:Sensitivity_Resolution}
  \end{subfigure}
  \caption{Global access $A/D$ when each input is varied around the default configuration (circled). The capacity-only access $A_\text{C}/D$ and geography-only access $A_\text{G}/D$ are shown where the input changes them.}
  \label{fig:Sensitivity}
\end{figure}

The results in \autoref{sec:Results} assume a machine capacity of 450 patients per linac per year, a Gaussian decay with a \qty{2}{h} driving-time limit, and capped 3SFCA allocation. \autoref{fig:Sensitivity} shows the sensitivity of global access to each of these inputs, varied independently around the default configuration. The exponential model uses a decay factor $\lambda = \qty{60}{min}$ and the power law $\beta = 1$ with a plateau $x_0 = \qty{10}{min}$, both cut off at \qty{2}{h}.

\autoref{fig:Sensitivity_Time} shows that access varies from \qty{28.0}{\%} of demand at \qty{30}{min} to \qty{60.0}{\%} at \qty{120}{min}, \autoref{fig:Sensitivity_Capacity} from \qty{42.3}{\%} at 250 patients per linac per year to \qty{64.2}{\%} at 550. \autoref{fig:Sensitivity_Decay} shows the choice of decay model can change global access by 6.1 points, from \qty{53.9}{\%} for the power law, whose impedance declines steeply beyond a short plateau, to \qty{62.4}{\%} for the step function, while \autoref{fig:Sensitivity_Allocation} shows variation by only 1 point. \autoref{fig:Sensitivity_Resolution} shows that global access varies by at most 0.3 points between H3 Resolution-3 and Resolution-5.

\clearpage
\section{Access Maps}~\label{app:Maps}

This appendix maps one country from each of the four categories of \autoref{sec:Countries}: strong access (United States), capacity constrained (Japan), geography constrained (Mexico), and doubly constrained (India).

All maps are shown at H3 Resolution-4 (average cell area of \qty{1770}{km^2}, centroid spacing of \qty{45}{km}), with demand and access calculated at Resolution-5 and summed into their parent hexagons. Demand and deficit share a common logarithmic colour scale across all four countries, from 1 to 50,000 patients per year per hexagon, so the maps can be compared directly.

\subsection{United States}
\begin{figure}[tbh!]
  \centering
  \begin{subfigure}{0.49\linewidth}
    \centering
    \includegraphics[width=\linewidth,height=0.38\textheight,keepaspectratio]{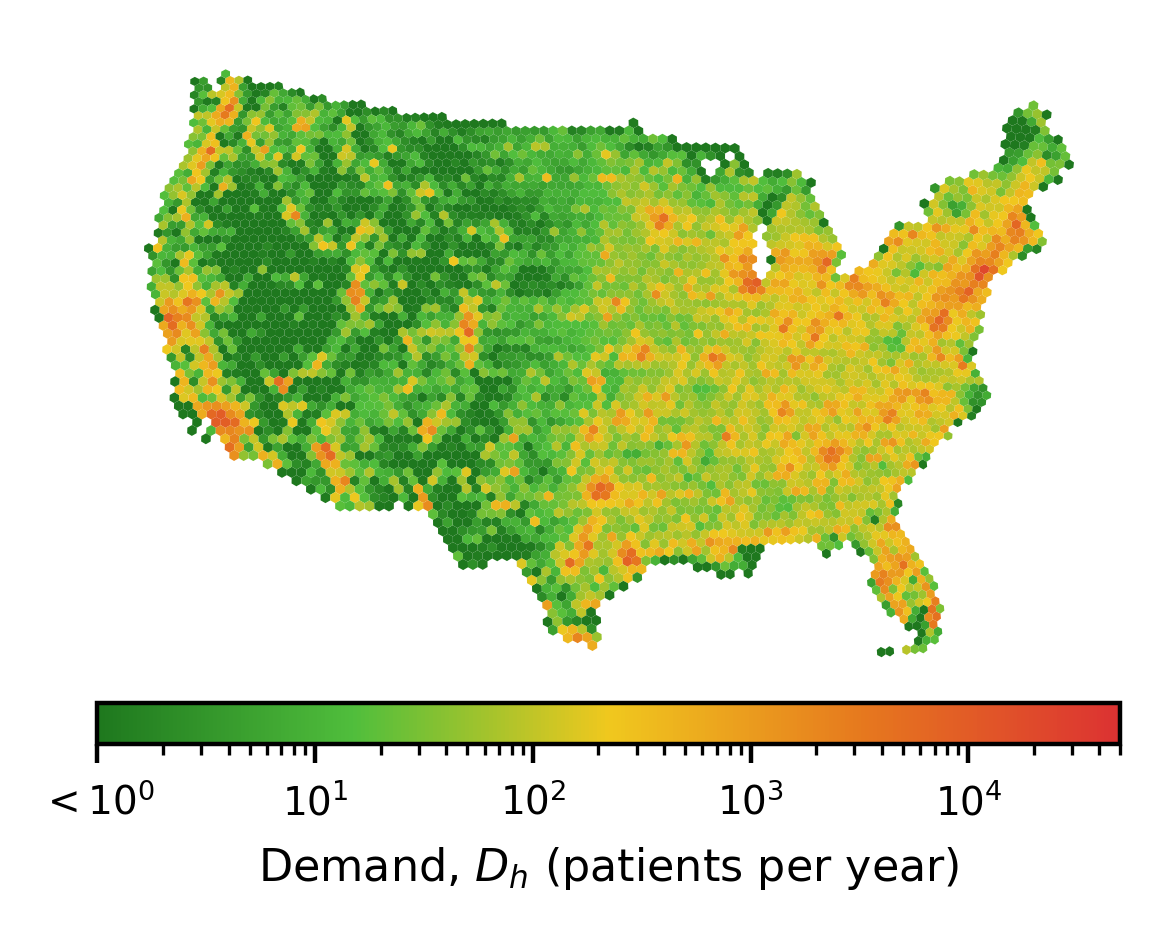}
    \caption{Demand, $D_h$.}
    \label{fig:USA_Demand}
  \end{subfigure}\hfill
  \begin{subfigure}{0.49\linewidth}
    \centering
    \includegraphics[width=\linewidth,height=0.38\textheight,keepaspectratio]{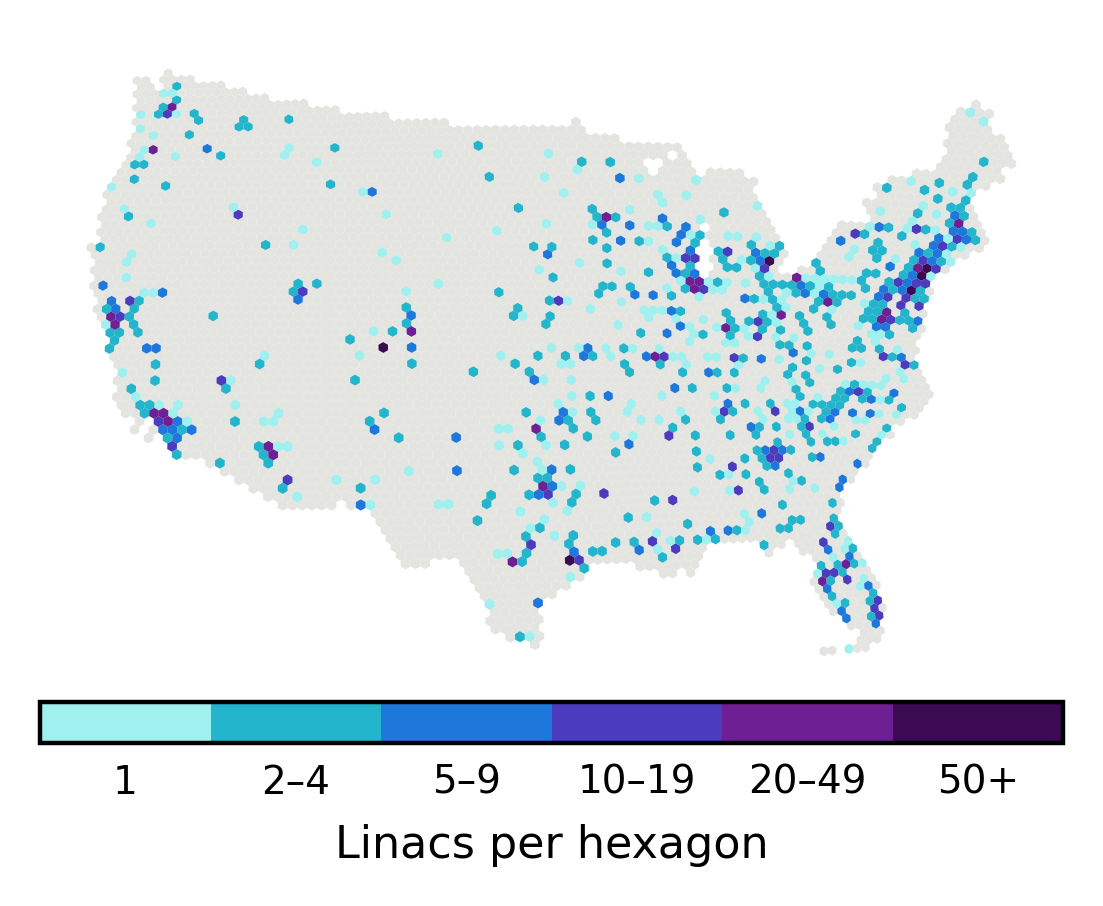}
    \caption{Linacs per hexagon.}
    \label{fig:USA_Mvm}
  \end{subfigure}\par\medskip
  \begin{subfigure}{0.49\linewidth}
    \centering
    \includegraphics[width=\linewidth,height=0.38\textheight,keepaspectratio]{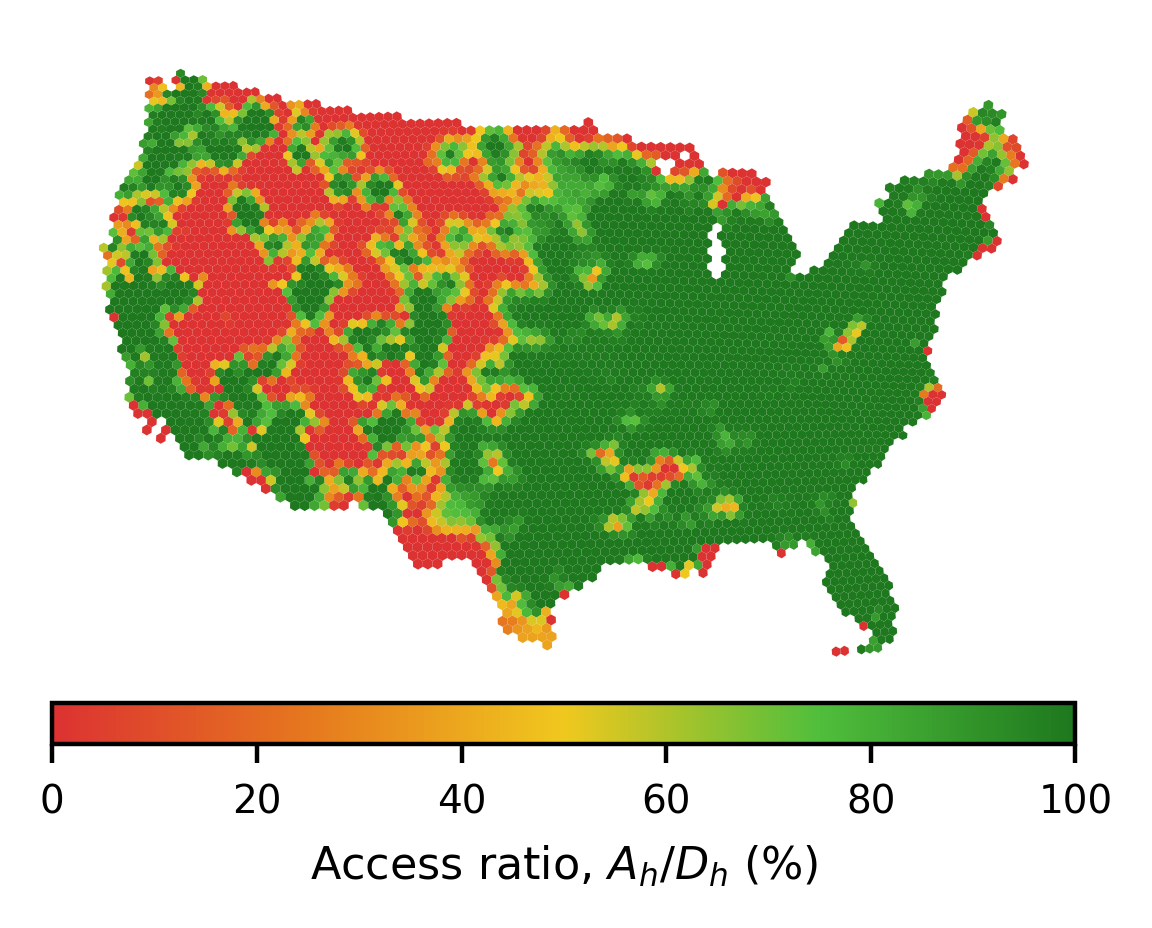}
    \caption{Access ratio, $A_h/D_h$.}
    \label{fig:USA_Ratio}
  \end{subfigure}\hfill
  \begin{subfigure}{0.49\linewidth}
    \centering
    \includegraphics[width=\linewidth,height=0.38\textheight,keepaspectratio]{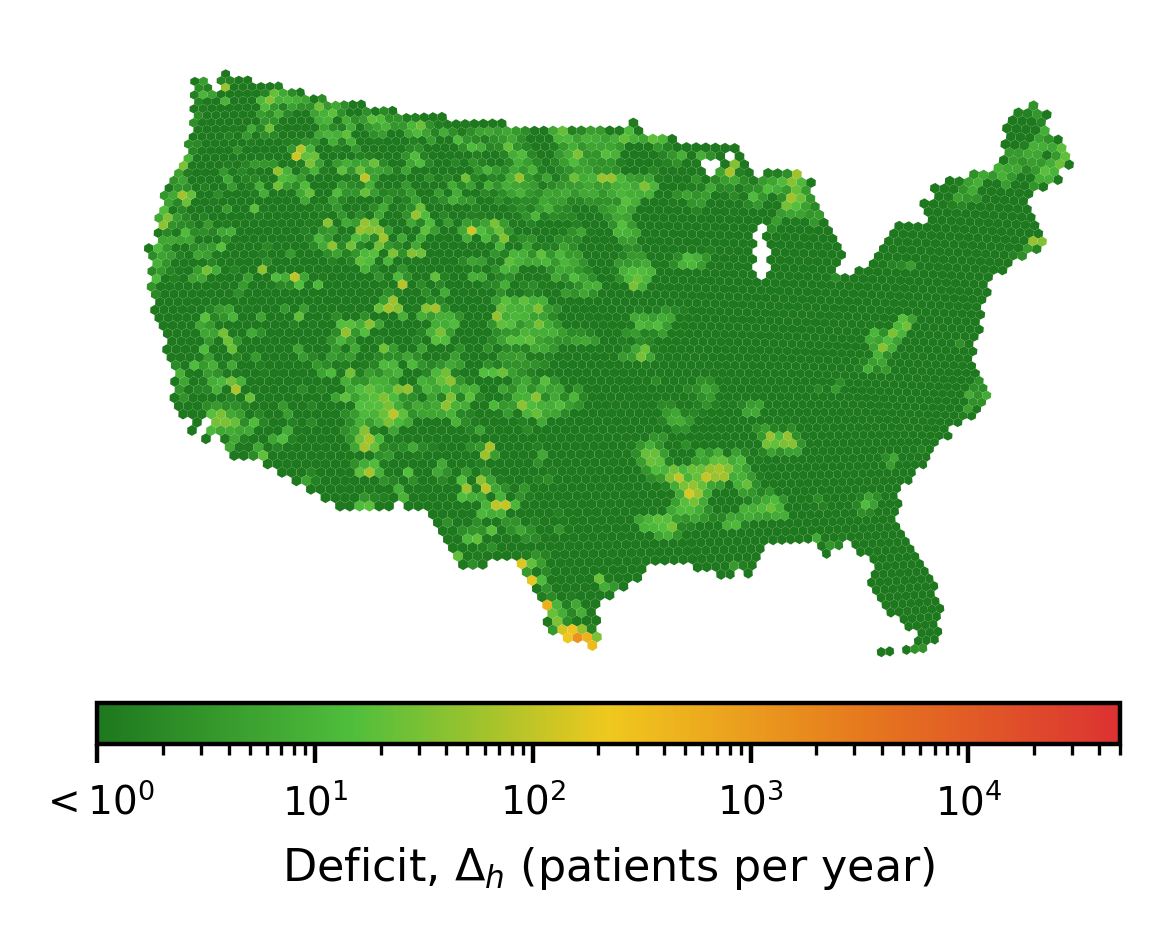}
    \caption{Deficit, $\Delta_h$.}
    \label{fig:USA_Deficit}
  \end{subfigure}
  \caption{Radiotherapy demand, linacs, access ratio and deficit across the contiguous United States (strong access) per H3 Resolution-4 hexagon. Alaska, Hawaii and the overseas territories are included in the calculation but not shown.}
  \label{fig:USA}
\end{figure}

\clearpage \subsection{Japan}
\begin{figure}[tbh!]
  \centering
  \begin{subfigure}{0.49\linewidth}
    \centering
    \includegraphics[width=\linewidth,height=0.38\textheight,keepaspectratio]{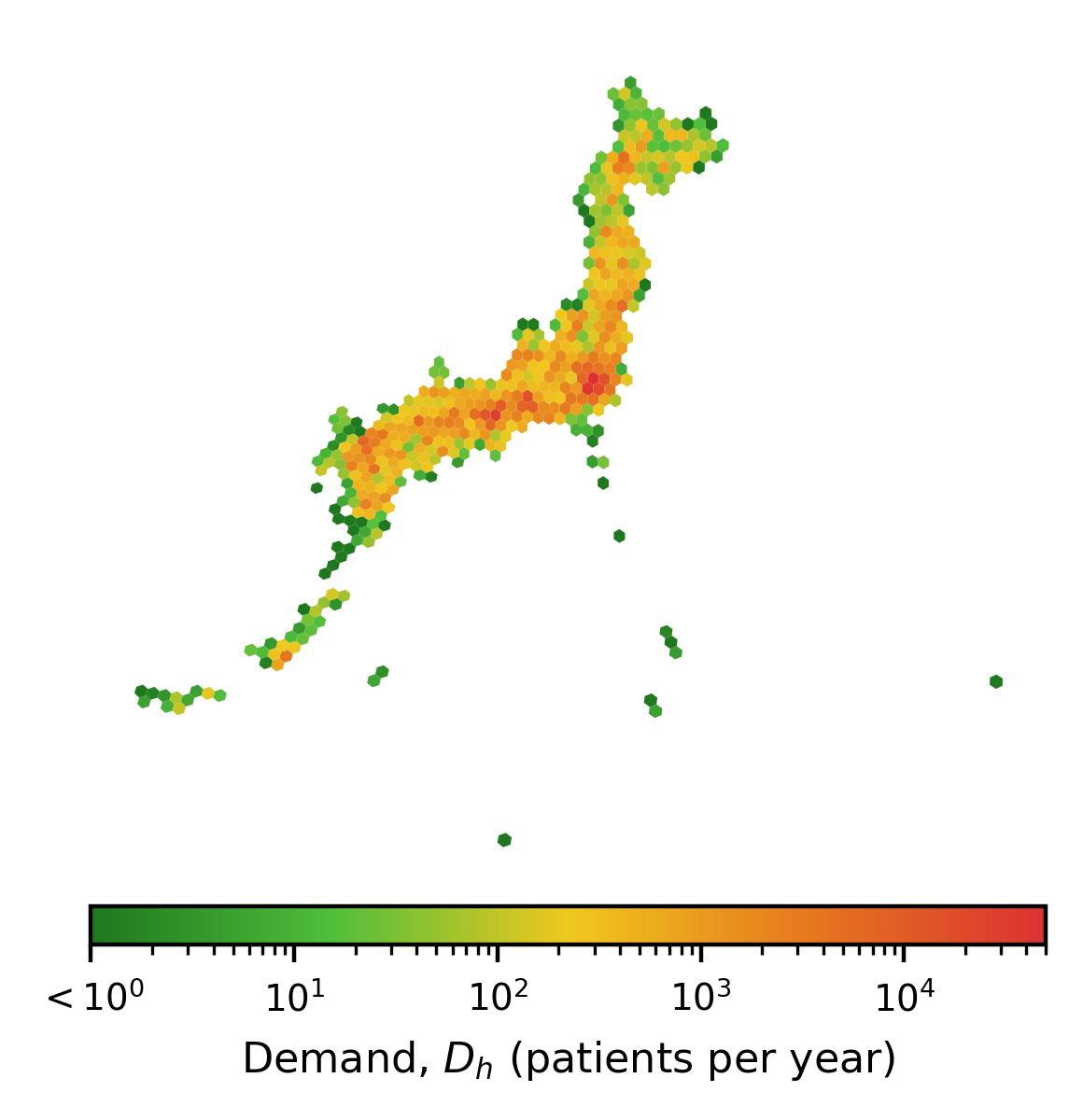}
    \caption{Demand, $D_h$.}
    \label{fig:JPN_Demand}
  \end{subfigure}\hfill
  \begin{subfigure}{0.49\linewidth}
    \centering
    \includegraphics[width=\linewidth,height=0.38\textheight,keepaspectratio]{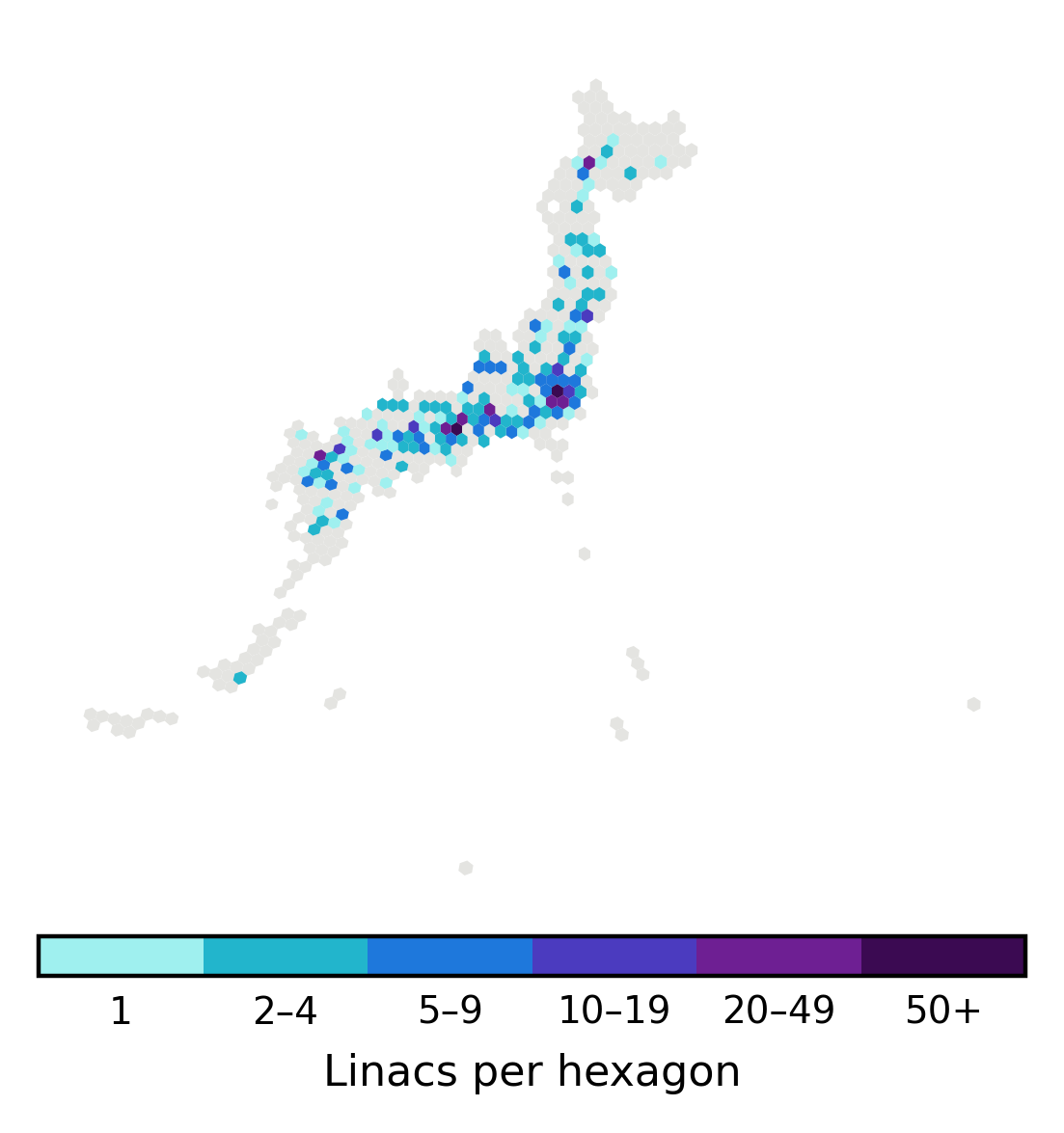}
    \caption{Linacs per hexagon.}
    \label{fig:JPN_Mvm}
  \end{subfigure}\par\medskip
  \begin{subfigure}{0.49\linewidth}
    \centering
    \includegraphics[width=\linewidth,height=0.38\textheight,keepaspectratio]{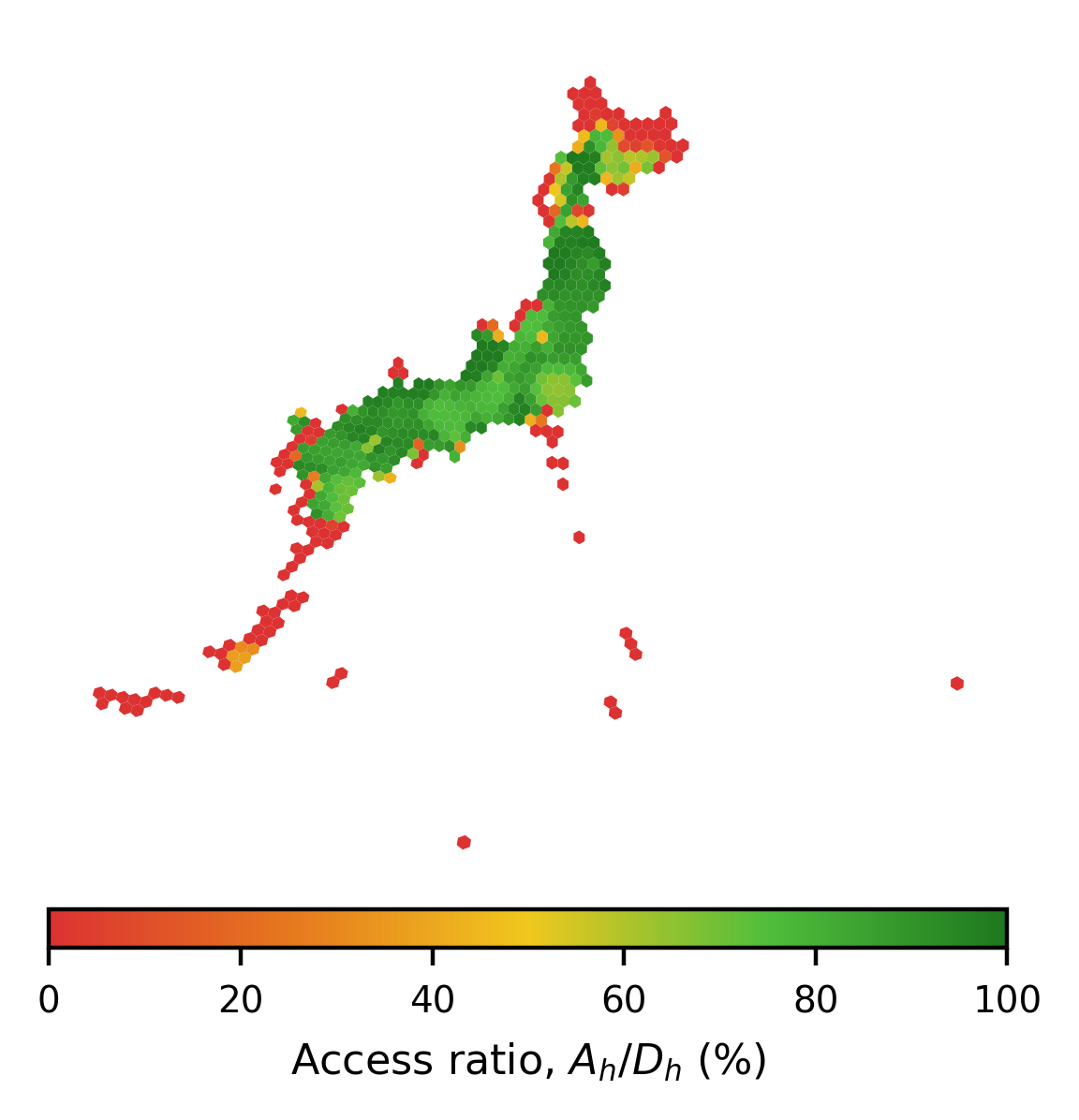}
    \caption{Access ratio, $A_h/D_h$.}
    \label{fig:JPN_Ratio}
  \end{subfigure}\hfill
  \begin{subfigure}{0.49\linewidth}
    \centering
    \includegraphics[width=\linewidth,height=0.38\textheight,keepaspectratio]{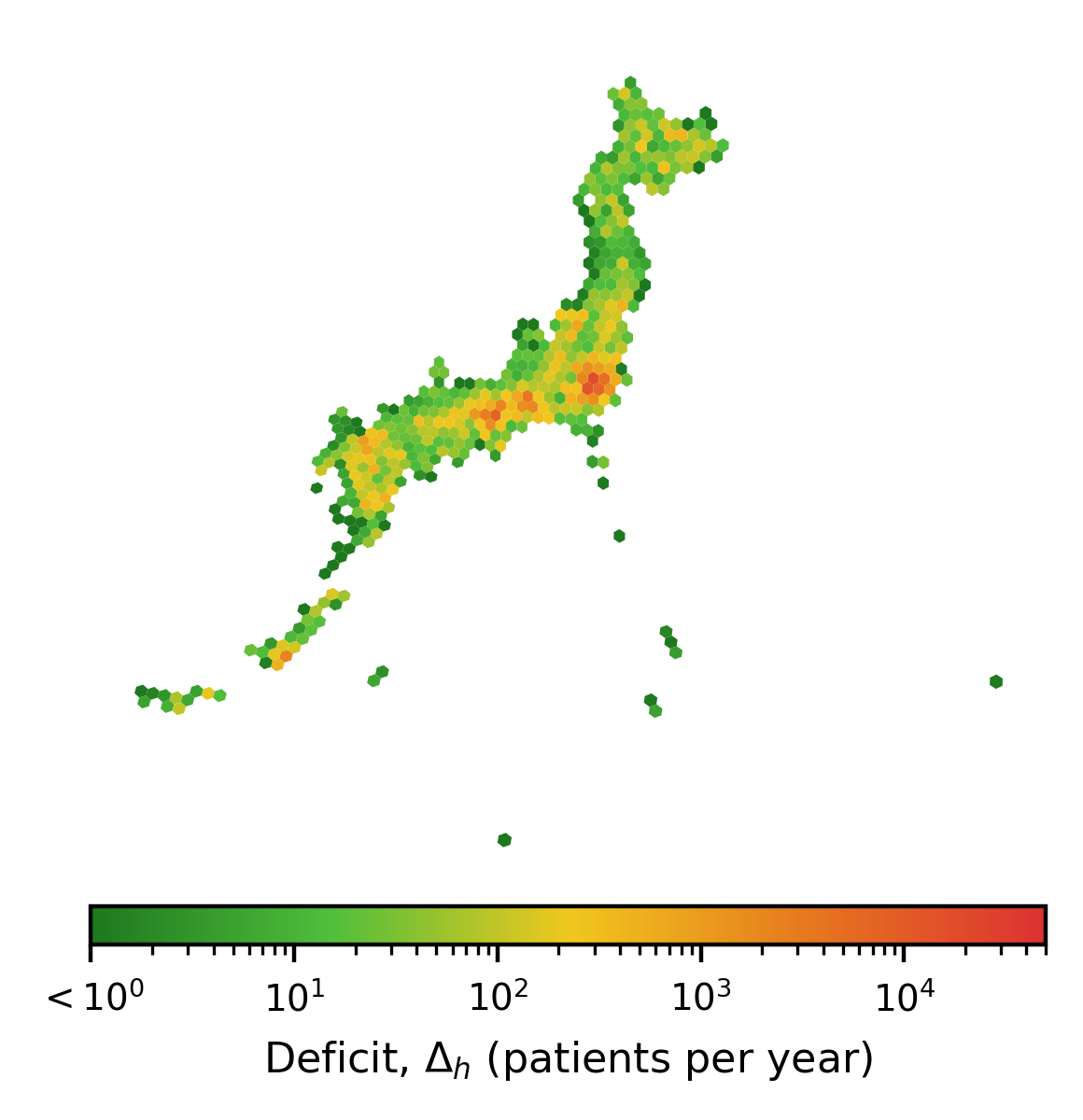}
    \caption{Deficit, $\Delta_h$.}
    \label{fig:JPN_Deficit}
  \end{subfigure}
  \caption{Radiotherapy demand, linacs, access ratio and deficit across Japan (capacity constrained) per H3 Resolution-4 hexagon.}
  \label{fig:JPN}
\end{figure}

\clearpage \subsection{Mexico}
\begin{figure}[tbh!]
  \centering
  \begin{subfigure}{0.49\linewidth}
    \centering
    \includegraphics[width=\linewidth,height=0.38\textheight,keepaspectratio]{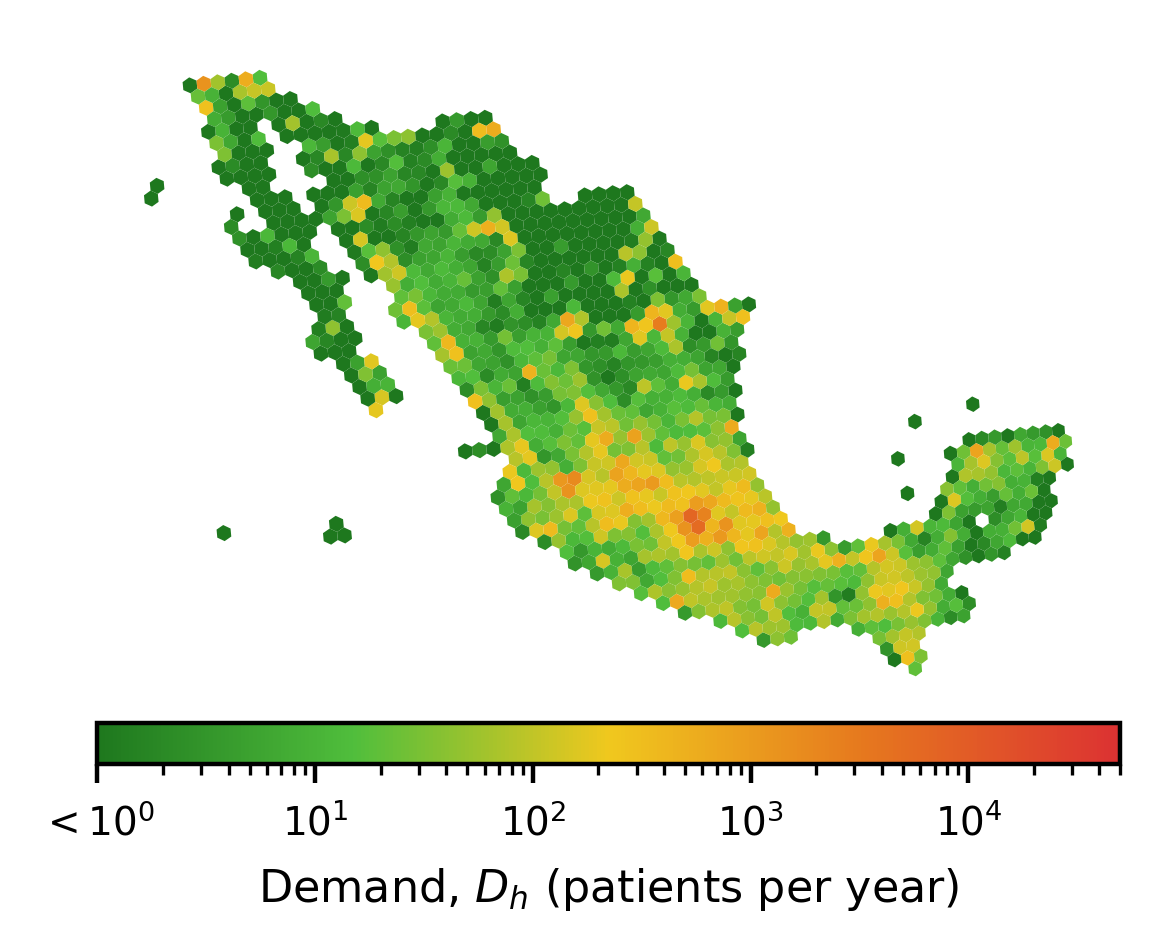}
    \caption{Demand, $D_h$.}
    \label{fig:MEX_Demand}
  \end{subfigure}\hfill
  \begin{subfigure}{0.49\linewidth}
    \centering
    \includegraphics[width=\linewidth,height=0.38\textheight,keepaspectratio]{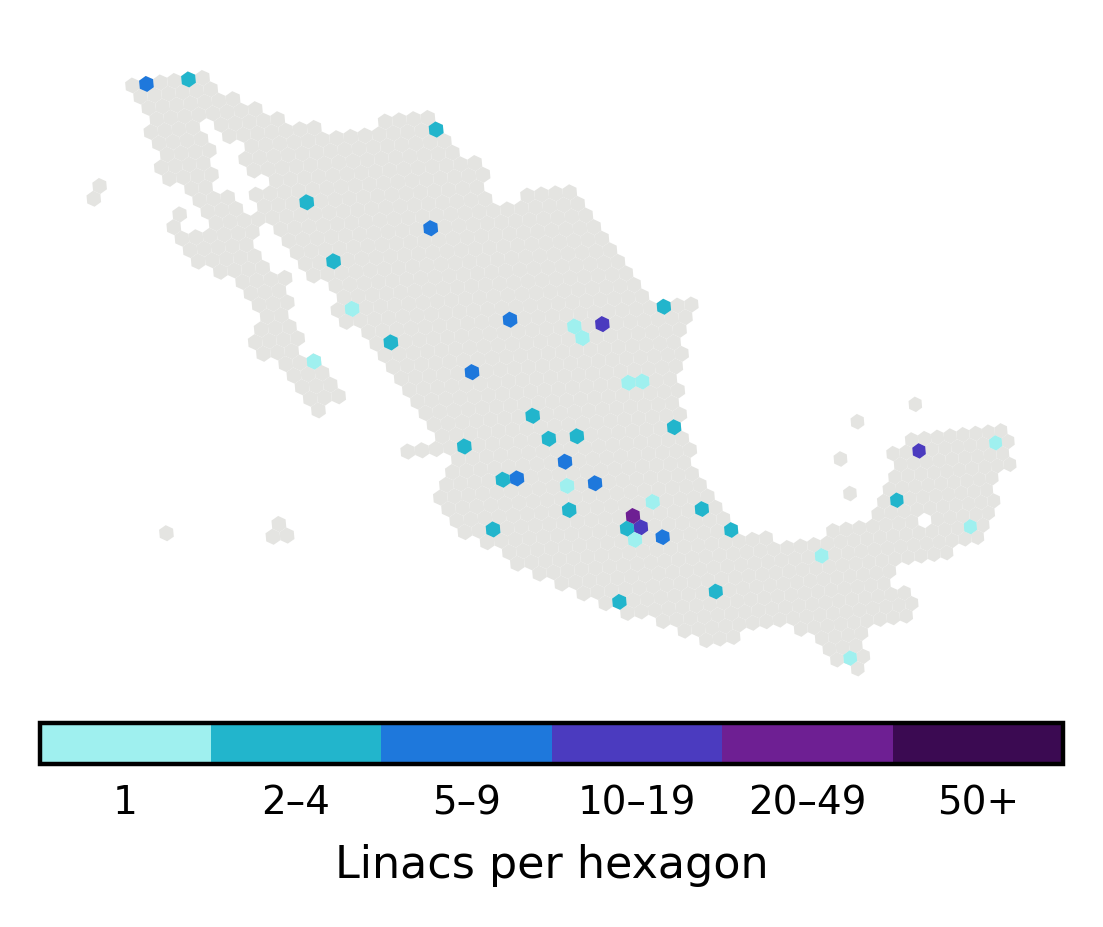}
    \caption{Linacs per hexagon.}
    \label{fig:MEX_Mvm}
  \end{subfigure}\par\medskip
  \begin{subfigure}{0.49\linewidth}
    \centering
    \includegraphics[width=\linewidth,height=0.38\textheight,keepaspectratio]{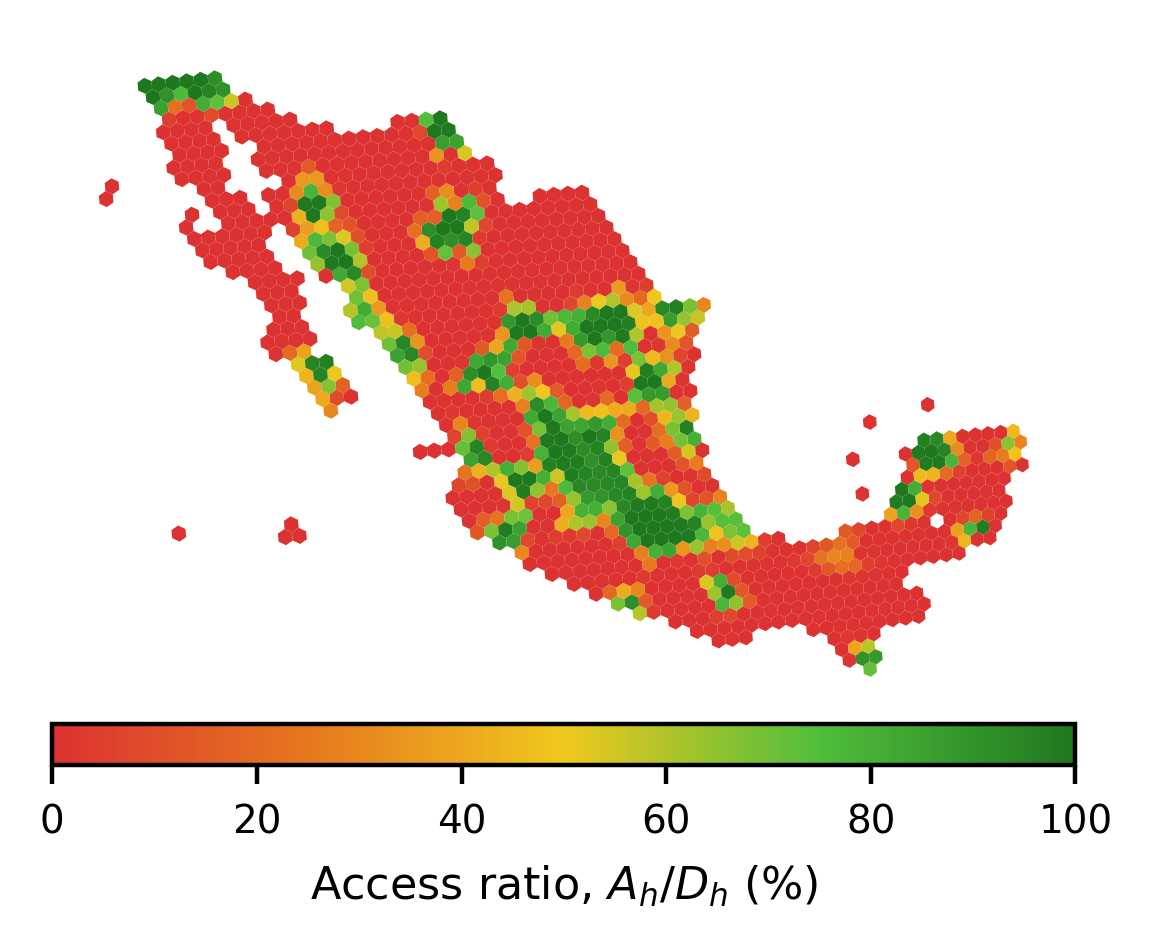}
    \caption{Access ratio, $A_h/D_h$.}
    \label{fig:MEX_Ratio}
  \end{subfigure}\hfill
  \begin{subfigure}{0.49\linewidth}
    \centering
    \includegraphics[width=\linewidth,height=0.38\textheight,keepaspectratio]{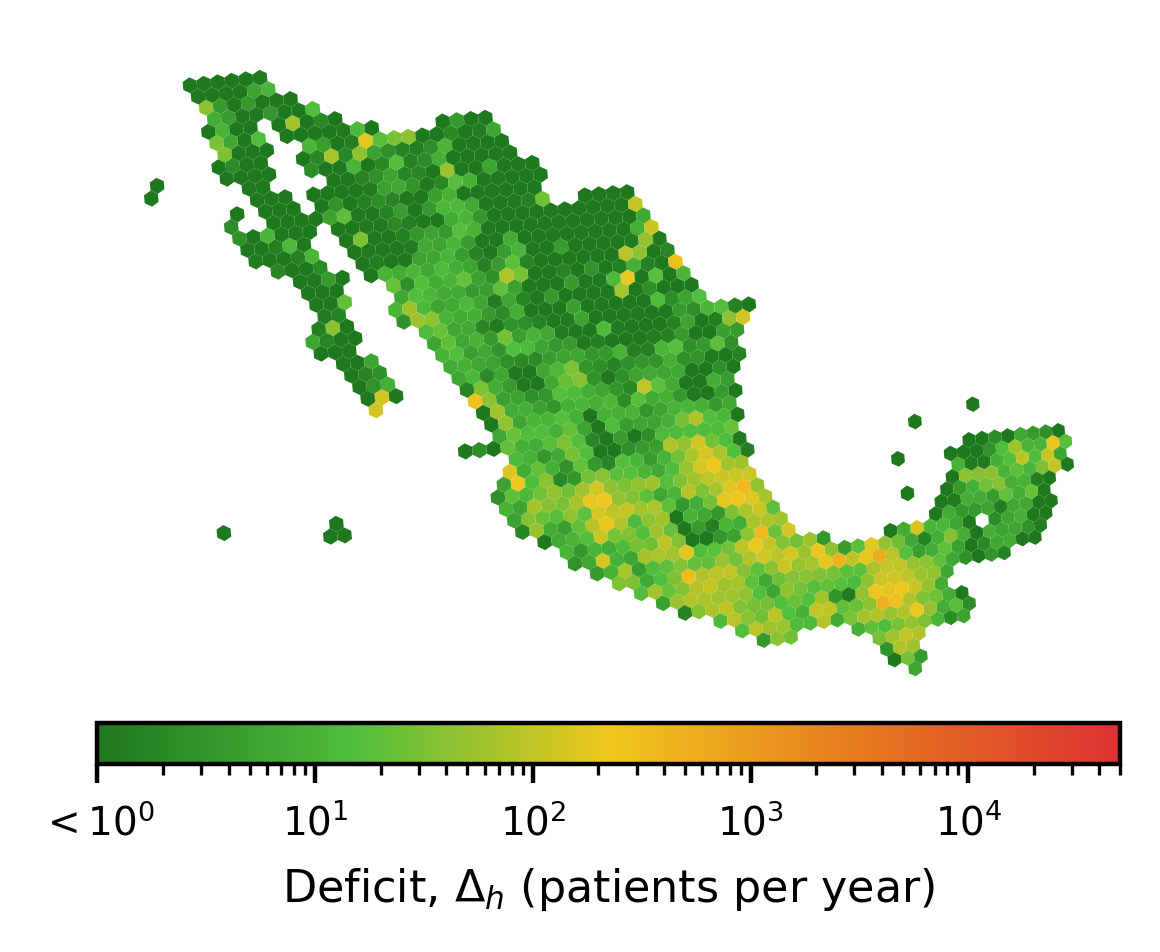}
    \caption{Deficit, $\Delta_h$.}
    \label{fig:MEX_Deficit}
  \end{subfigure}
  \caption{Radiotherapy demand, linacs, access ratio and deficit across Mexico (geography constrained) per H3 Resolution-4 hexagon.}
  \label{fig:MEX}
\end{figure}

\clearpage \subsection{India}
\begin{figure}[tbh!]
  \centering
  \begin{subfigure}{0.49\linewidth}
    \centering
    \includegraphics[width=\linewidth,height=0.38\textheight,keepaspectratio]{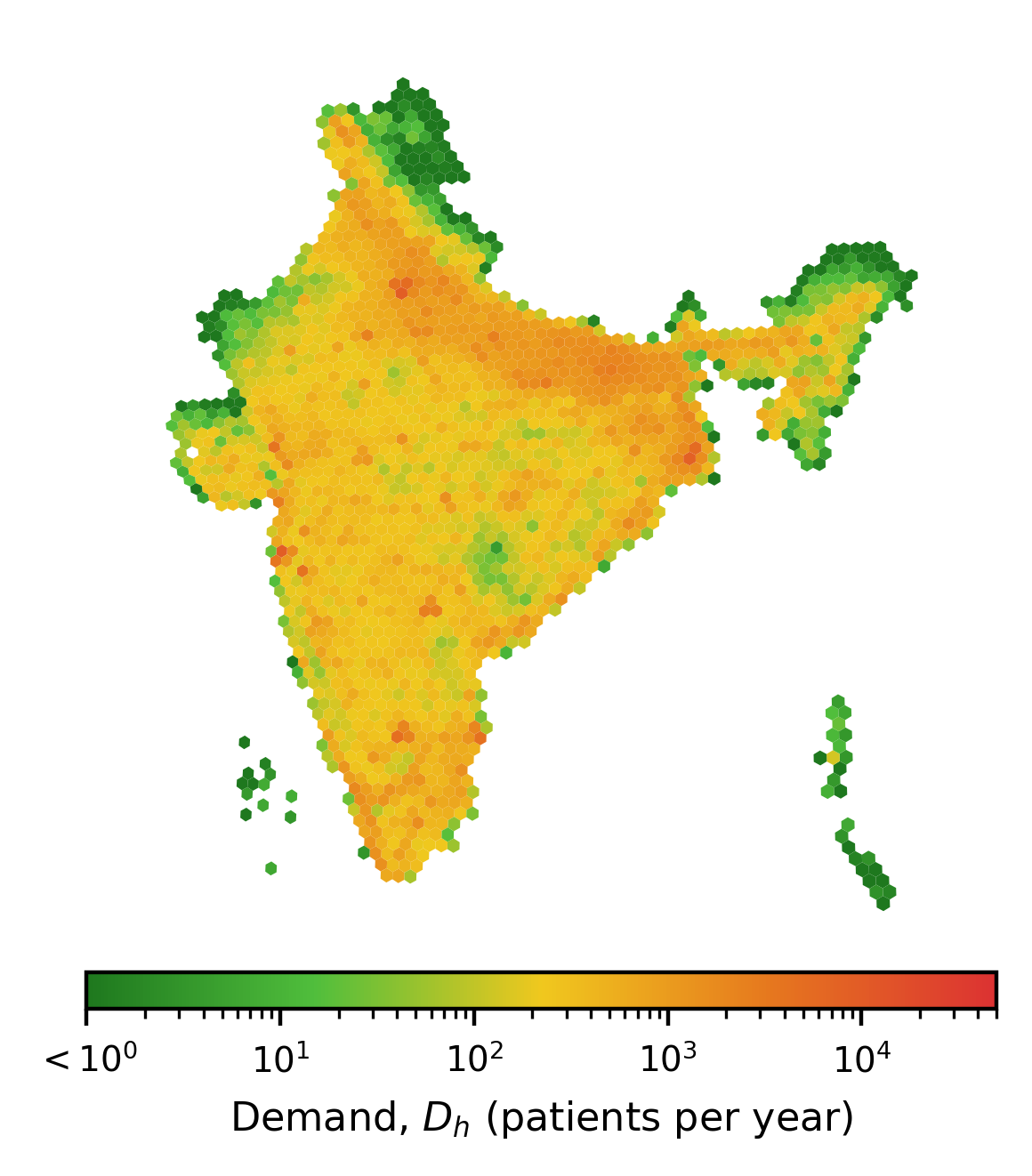}
    \caption{Demand, $D_h$.}
    \label{fig:IND_Demand}
  \end{subfigure}\hfill
  \begin{subfigure}{0.49\linewidth}
    \centering
    \includegraphics[width=\linewidth,height=0.38\textheight,keepaspectratio]{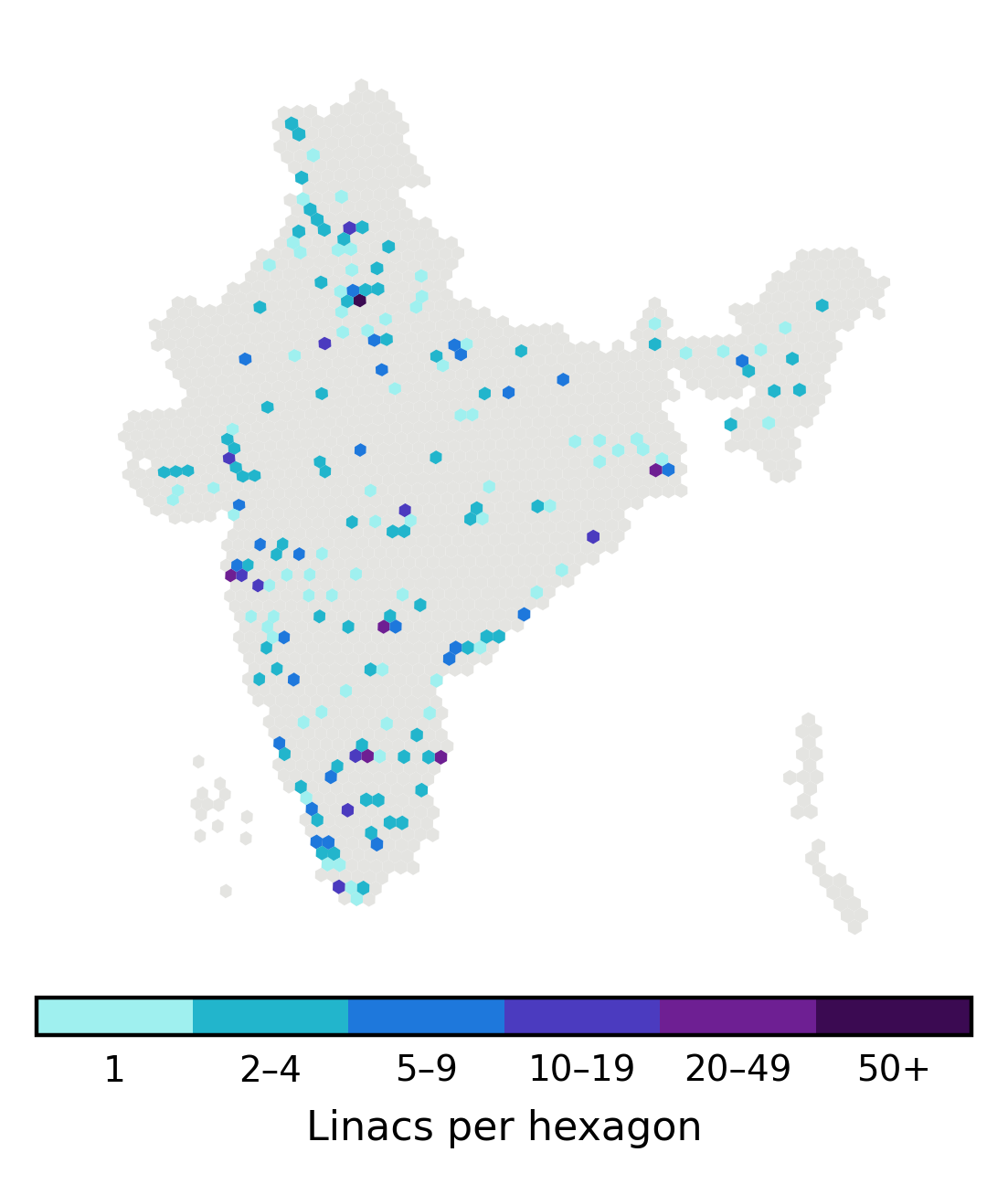}
    \caption{Linacs per hexagon.}
    \label{fig:IND_Mvm}
  \end{subfigure}\par\medskip
  \begin{subfigure}{0.49\linewidth}
    \centering
    \includegraphics[width=\linewidth,height=0.38\textheight,keepaspectratio]{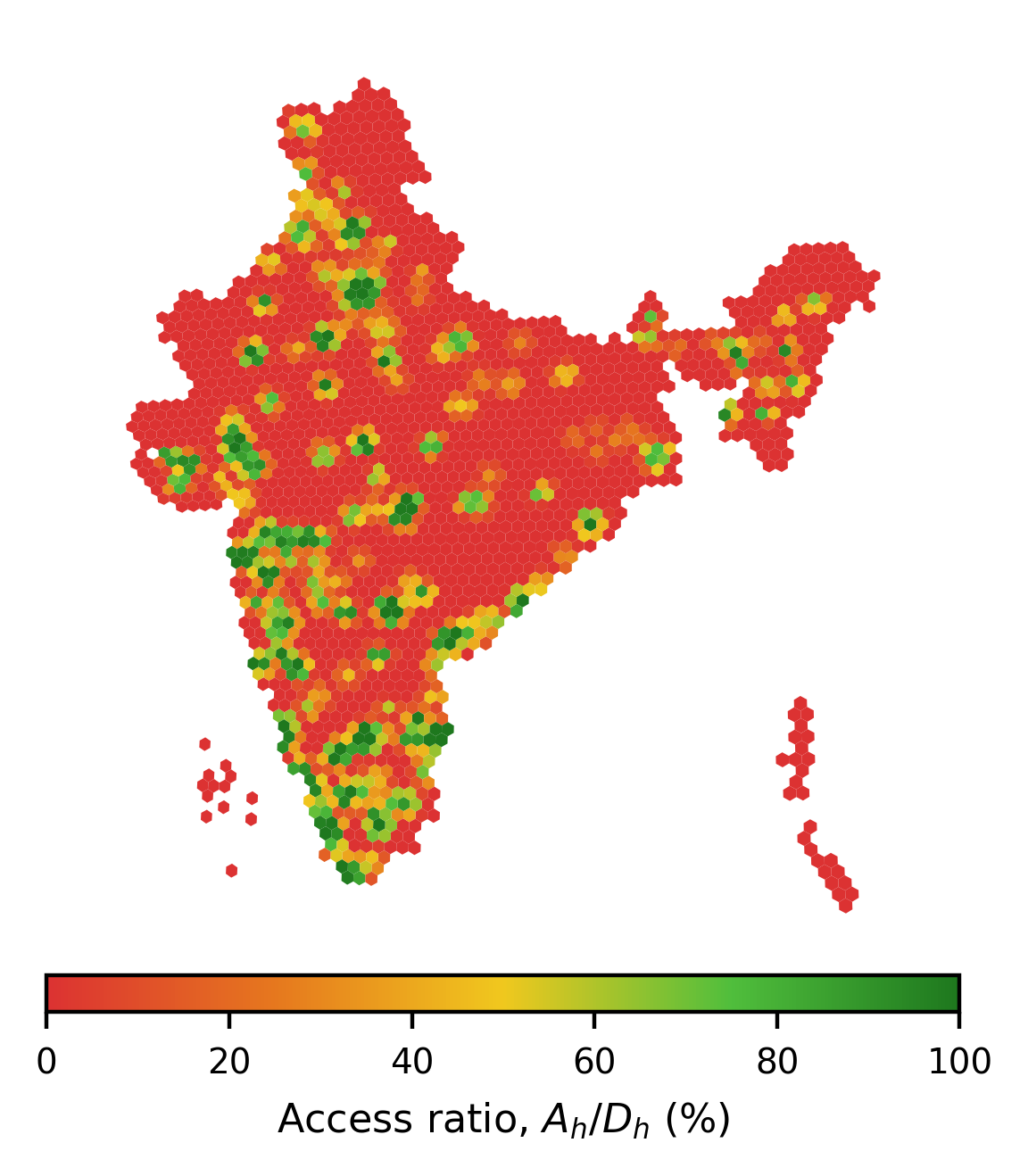}
    \caption{Access ratio, $A_h/D_h$.}
    \label{fig:IND_Ratio}
  \end{subfigure}\hfill
  \begin{subfigure}{0.49\linewidth}
    \centering
    \includegraphics[width=\linewidth,height=0.38\textheight,keepaspectratio]{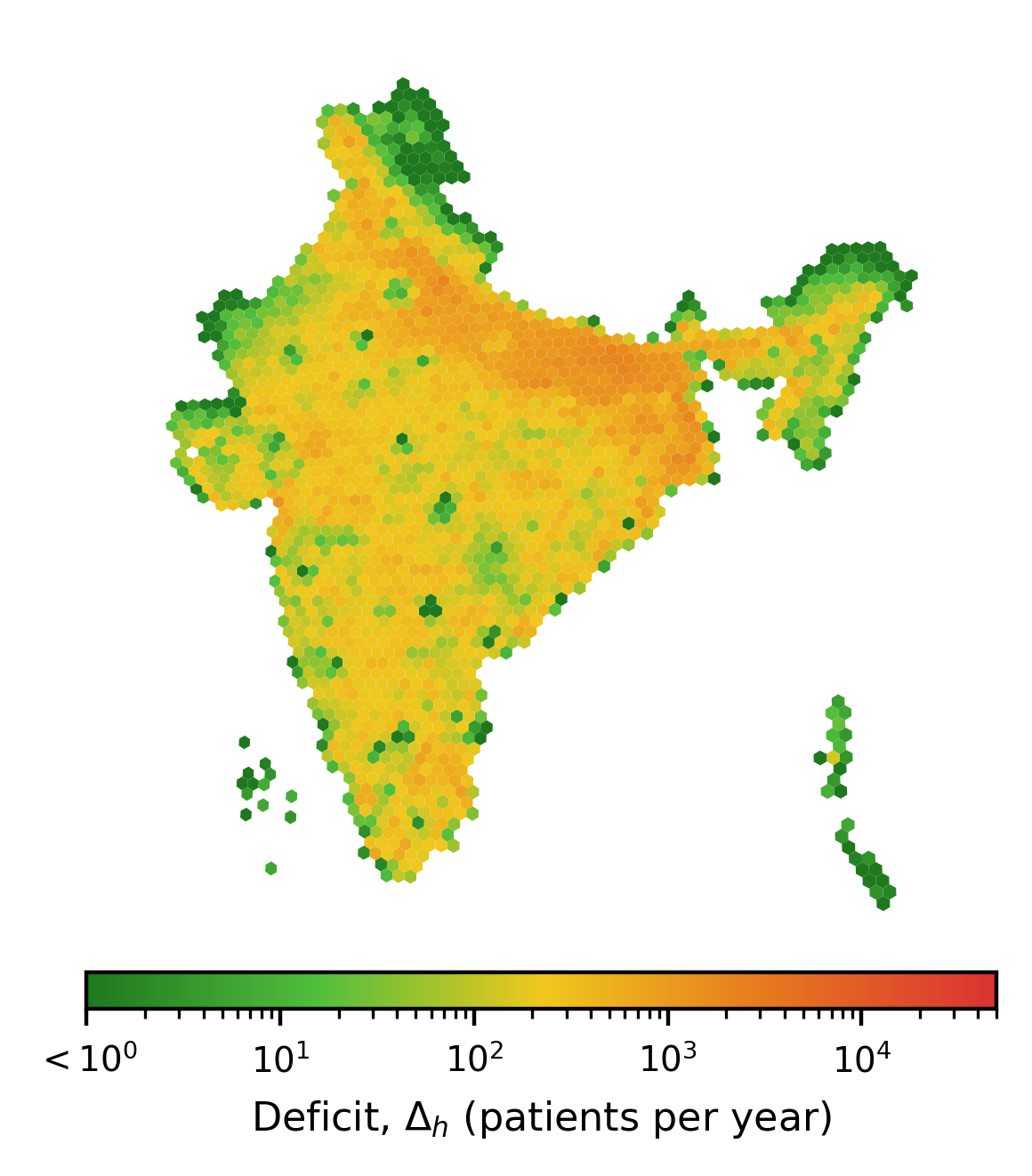}
    \caption{Deficit, $\Delta_h$.}
    \label{fig:IND_Deficit}
  \end{subfigure}
  \caption{Radiotherapy demand, linacs, access ratio and deficit across India (doubly constrained) per H3 Resolution-4 hexagon.}
  \label{fig:IND}
\end{figure}

\end{document}